\documentclass[aps,pra,twocolumn,superscriptaddress,floatfix]{revtex4-1}
\usepackage[svgnames]{xcolor}
\usepackage[colorlinks=true,urlcolor = teal, citecolor = blue, linkcolor= magenta, bookmarks=false]{hyperref}
\usepackage{amsfonts,amsmath,amssymb}
\usepackage{graphicx}
\usepackage{dsfont}

\usepackage{tikz}
\usetikzlibrary{arrows,topaths,shapes.geometric}
\usepackage{pgfplots}

\def\ii{{\rm i}}
\newcommand{\dd}{{\rm d}}

\def\expect#1{\langle#1\rangle}

\def\ol#1{\bar{#1}}

\def\ua{{\uparrow}}
\def\da{{\downarrow}}

\def\hstar{\,\hat{\star}\,}
\def\rhoh{\ol{\rho}}
\newcommand{\uu}{{\mathfrak{u}}}

\newcommand{\calA}{\mathcal{A}}
\newcommand{\calC}{\mathcal{C}}
\newcommand{\calD}{\mathcal{D}}

\newcommand{\calO}{\mathcal{O}}

\newcommand{\calS}{\mathcal{S}}

\newcommand{\bq}{\mathbf{q}}
\newcommand{\bj}{\mathbf{j}}

\newcommand{\bY}{\mathbf{Y}}

\newcommand{\p}{k}
\newcommand{\pdr}{p}

\newcommand{\e}{e}
\newcommand{\edr}{\varepsilon}
\newcommand{\be}{\mathbf{e}}

\begin{document}

\title{Ballistic transport in the one-dimensional Hubbard model:\\ the hydrodynamic approach}

\author{Enej Ilievski}
\affiliation{Institute for Theoretical Physics Amsterdam and Delta Institute for Theoretical Physics,
University of Amsterdam, Science Park 904, 1098 XH Amsterdam, The Netherlands}

\author{Jacopo De Nardis}
\affiliation{D\'epartement de Physique, Ecole Normale Sup\'erieure,
PSL Research University, CNRS, 24 rue Lhomond, 75005 Paris, France}

\date{\today}

\begin{abstract}
We outline a general formalism of hydrodynamics for quantum systems with multiple particle species which undergo
completely elastic scattering. In the thermodynamic limit, the complete kinematic data of the problem consists of the particle
content, the dispersion relations, and a universal dressing transformation which accounts for interparticle interactions.
We consider quantum integrable models and we focus on the one-dimensional fermionic Hubbard model.
By linearizing hydrodynamic equations, we provide exact closed-form expressions for Drude weights, generalized static charge 
susceptibilities and charge-current correlators valid on hydrodynamic scale, represented as integral kernels operating
diagonally in the space of mode numbers of thermodynamic excitations. We find that, on hydrodynamic scales, Drude weights manifestly 
display Onsager reciprocal relations even for generic (i.e. non-canonical) equilibrium states, and establish a generalized detailed 
balance condition for a general quantum integrable model. We present the first exact analytic expressions for the general Drude 
weights in the Hubbard model, and explain how to reconcile different approaches for computing Drude weights from the previous 
literature.
\end{abstract}


\maketitle


In past few years, a lot of interest has been devoted to studying various paradigms of non-ergodic many-body physics,
such as quantum quenches, equilibration to generalized Gibbs ensembles and phenomenon of
pre-thermalization \cite{Polkovnikov_review,Calabrese_intro,Gogolin}.
One of the prominent recent results is the formalism of generalized hydrodynamics developed in \cite{BCNF16,CDY16},
with a large number of subsequent studies investigating its various aspects and
applications~\cite{DoYo16,DoSY17,DoSp17,DoYC17,DDKY17,BVKM17-2,Piroli-longBertini,VAlba_EntropyTransport},
including the exact computation of Drude weights in the Heisenberg model XXZ spin-1/2 chain \cite{ID17}.
In analogy to the conventional theory of hydrodynamics~\cite{Spohn_book}, the authors of \cite{DS17} just recently obtained
a closed formula for Drude weights expressed in terms of local equilibrium state functions for the case of integrable Bose
gas (Lieb--Liniger model) and conjectured that similar formulae may hold in quantum integrable models more generally. In this work, we 
go a step further and extend the formalism to integrable models which possess physical particles with internal degrees of freedom and 
are solvable by \emph{nested Bethe Ansatz}.
Nesting is referred to the situation when physical degrees of freedom are associated with a higher rank symmetry group, leading to 
eigenfunctions with a hierarchical structure of internal quantum numbers and elementary excitations of different flavours.
While studies of such models has been traditionally focused
on Gibbs equilibrium~\cite{EKS91,EKS92,EK94,Hubbard_book,AdSCFT_review,FQ12}, they have also been recently studied
in the nonequilibrium context \cite{CalabreseNested,RCK16}.

The chief aspect in which interacting quantum integrable theories differ from widely studied noninteracting systems is
the dressing of (quasi)particle excitations, i.e. a process in which bare properties of the particle-hole type of excitations
renormalize in the presence of interactions with a non-trivial reference (vacuum) state. The task of classifying excitations has
been traditionally restricted to ground states for some of the simplest Bethe Ansatz solvable models \cite{Korepin_book}, and
subsequently extended to some important examples of exactly solvable models of correlated
electrons \cite{Bares91,EK94,Hubbard_book,QF13}. A comprehensive exposition of the dressing formalism for grand canonical ensembles in 
nested Bethe Ansatz models can be found in \cite{QF13}.

\paragraph*{Dressing formalism.}
Integrable theories exhibit a completely elastic (factorizable) scattering of particle-like excitations~\cite{ZZ79}. Properties
of such excitations represent the \emph{kinematic data} of the theory. In particular, in Bethe Ansatz solvable models (see 
e.g.~\cite{Korepin_book,Hubbard_book}) thermodynamic excitations relative to a bare vacuum \footnote{In fermionic interacting 
integrable models there exist distinct inequivalent possibilities of choosing a bare vacuum.
Despite this results in different sets of excitations, various choices have no effect on the values of physical
observables.} can be inferred from the solutions to (nested) Bethe equations. The latter in a finite volume take the form
$e^{\ii p_{\alpha}(u^{(\alpha)}_{k})}\prod_{\beta}\prod_{j=1}^{N_{\beta}}S_{\alpha \beta}(u^{(\alpha)}_{k},u^{(\beta)}_{j}) = 1$,
imposing single-valuedness of many-body eigenstates.
Here the sets of quantum numbers $\{u^{(\alpha)}_{k}\}$ are called the Bethe roots and represent rapidity variables for
distinct species (or flavours) of elementary excitations. The number and types of excitations depends on the model and can be
inferred with aid of representation theory of the underlying quantized Lie (super)algebra.
Elementary excitations typically form complexes which are interpreted as bound states.
The emergent thermodynamic particle content, which can be inferred by e.g. analysing the $L\to \infty$ limit of Bethe equations,
is generally different from elementary excitations and is labeled by a pair of \emph{mode numbers},
a particle type index $a$ and a real rapidity variable $u$. The complete kinematic data
are obtained from the bare momenta $\p_{a}(u)$ and energies $\e_{a}(u)$, and interparticle scattering phase
shifts $\phi_{ab}(u,w)$. Once given these functions, no explicit operator representation of the Hamiltonian and its conservation
laws is ever required. In this work we present the details of the entire formalism for the non-trivial case of the (fermionic) 
Hubbard model. \\

A distinguished feature of integrable systems is a macroscopic number of local conservation laws which can be formally
expressed in terms of a discrete basis of local charges $Q_{i}=\sum_x \,q_{i}(x)$, with $x$ labelling lattice sites.
The associated currents $J_{i}=\sum_x \,j_{i}(x)$ are defined with aid of the continuity equation,
$\partial_{t}\hat{Q}_{i}+\partial_{x}\hat{J}_{i}=0$.
The key concept of the hydrodynamic approach is the dressing of bare energies $\e_{a}\mapsto \edr_{a}$ and momenta
$\p_{a}\mapsto \pdr_{a}$ of particle excitations, which can be presented in a compact form
\begin{equation}
\edr^{\prime}_{a} = \Omega_{ab}\star \e^{\prime}_{b},\quad \pdr^{\prime}_{a} = \Omega_{ab}\star \p^{\prime}_{b}.
\label{eqn:dressed_derivative_energy_momentum}
\end{equation}
with convolution $(\Omega_{ab}\star f_{b})(u)=\sum_{b}\int \dd w\,\Omega_{ab}(u,w)f_{b}(w)$.
In interacting quantum integrable models solvable by (nested) Bethe Ansatz, the matrix convolution kernel $\Omega$ takes a
\emph{universal form}
\begin{equation}
\left(\Omega^{-1}\right)_{ab}(u,w) = \delta_{ab}\delta(u-w) + K_{ab}(u-w)\vartheta_{b}(w)\sigma_{b}.
\label{eqn:dressing_transformation}
\end{equation}
with kernels $K_{ab}(u,w)$ defined as derivatives of the scattering phase shifts $\phi_{ab}(u,w)=\phi_{ab}(u-w)$,
$K_{ab}(u)=\tfrac{1}{2\pi \ii}\partial_{u}\phi_{ab}(u)$, and $\sigma_{a}={\rm sign}(\p^{\prime}_{a}(u))$.
The (Fermi) filling functions  $\vartheta_{a}(u)$  specify the fraction of occupied modes with rapidities inside a small interval 
around $u$.

Dispersion relations of excitations $\varepsilon_{a}(u)$ depend on a many-body vacuum which is uniquely specified
by the rapidity distributions $\rho_{a}(u)$.
In terms of (thermodynamic) particle excitations, the equilibrium averages of charge and current densities decompose as
$q_{i} = \sum_{a}\int \dd u\,q_{i,a}(u)\rho_{a}(u)$, $j_{i} = \sum_{a}\int \dd u\,q_{i,a}(u)j_{a}(u)$,
where $j_{a}(u)=\rho_{a}(u)v^{\rm dr}_{a}(u)$ are the current densities per mode~\cite{BCNF16,CDY16}.
The group velocities of propagating particles are thus state-dependent,
$v^{\rm dr}_{a}(u)=\edr^{\prime}_{a}(u)/\pdr^{\prime}_{a}(u)$.

We furthermore introduce the \emph{effective charges} as the bare charges renormalized under transformation $\Omega$, namely
the effective value of a local charge density $q_{i}$ is obtained as
\begin{equation}
q^{\rm eff}_{a,i} = \Omega_{ab}\star q_{b,i} =  \partial_{\mu_{i}}\log \left(\vartheta_{a}^{-1} - 1\right).
\label{eqn:dressing_abstract}
\end{equation} 
Here parameters $\mu_{i}$ are the chemical potentials of a local (generalized) equilibrium ensemble parametrized as
$\hat{\varrho}\simeq \exp{(-\sum_{i}\mu_{i}\hat{Q}_{i})}$~\cite{VR16_review,EF16_review,IlievskiGGE15}.
It is important to emphasize that despite the derivatives of dressed energies satisfying
$\edr^{\prime}_{a}=\Omega_{ab}\star\,\e^{\prime}_{b}=(\e^{\prime}_{a})^{\rm eff}$,
the effective charges are \emph{not} the proper dressed charges associated with an 
excitation, and specifically $\edr_{a}\neq \e_{a}^{\rm eff}$.
We moreover note that with aid of fusion identities among the scattering kernels,
the transformation \eqref{eqn:dressing_transformation} can be decoupled to a quasi-local form in the
mode space, cf. Supplemental Material~\cite{SM} (SM) for explicit form for Hubbard model.

\paragraph*{Drude weights.}
In this work, we shall mainly be concerned with general off-diagonal Drude weights
\begin{equation}
\calD^{(i,j)}=\frac{\beta}{2}\lim_{t\to \infty}\int_{\tau=0}^{t} \dd \tau\,\expect{\hat{J}_{i}(\tau)\,\hat{j}_{j}(0)}_{\rm c},
\label{eqn:Drude_ij}
\end{equation}
which represent magnitudes of the singular parts of the zero-frequency generalized conductivities~\cite{Kubo57,Mahan_book},
${\rm Re}\,\sigma_{ij}(\omega)=2\pi\,\calD^{(i,j)}\delta(\omega)+\sigma^{\rm reg}_{ij}(\omega)$. We use
$\expect{\cdot}_{\rm c}$ to denote the connected part of the equilibrium expectation values. Although we shall restrict
ourselves to grand canonical equilibria, our formalism applies (without modifications) to general local equilibrium states.

An exact representation for $\calD^{(i,j)}$ can be given in terms of the static covariance matrix $\calC$,
$\calC_{ij} = \expect{Q_{i}q_{j}}_{\rm c}$, with diagonal components $\chi_{i}=\calC_{ii}$ representing
(generalized) static susceptibilities, and charge-current correlators
(overlaps) $\calO$, $\calO_{ij} = \expect{Q_{i}j_{j}}_{\rm c}$. Explicit expressions in terms of thermodynamic state functions
can be found in \cite{SM}.
The time-averaged current--current correlator Eq.~\eqref{eqn:Drude_ij} can be projected onto the subspace formed by local
conserved quantities which yields the well-known Mazur--Suzuki equality~\cite{Mazur69,Suzuki71} and proves useful for
bounding dynamical susceptibilities~\cite{ZNP97}. In matrix notation the latter reads
$\calD^{(i,j)} = \tfrac{\beta}{2}\calO_{ik}\,(\calC^{-1})_{kl}\,\calO_{lj}$~\cite{IP12}.

A central result of our work is that on hydrodynamic scale, static charge-charge, charge-current correlations, and generic Drude 
weights, all assume a universal mode decomposition (writing formally $\calA\in \{\calC,\calD,\calO\}$)
\begin{equation}
\mathcal{A}_{ij} =
\sum_{a}\int \dd u\,q^{\rm eff}_{a,i}(u)\,A_{a}(u)\,q^{\rm eff}_{a,j}(u),
\label{eqn:resolution}
\end{equation}
which has exactly the same form as in the case of a single-component interacting integrable
Bose gas derived in a recent paper \cite{DS17}.
Importantly, in the above formula the kernels $A_{a}(u)$ and effective charges $q^{\rm eff}_{a,i}$ are expressible in terms
of properties of equilibrium states which can be efficiently computed within
Thermodynamic Bethe Ansatz (TBA) method~\cite{YY69,Takahashi71,Gaudin71}.
It is noteworthy that Eq.~\eqref{eqn:resolution} is written solely in the mode space, i.e. it acts (diagonally)
on particle labels and rapidities, and that no explicit knowledge of a complete set of local charges is ever required
in a computation. Indeed, thermodynamic expectation values of local charges are expressible as \emph{linear} functionals of particles' 
rapidity distributions (see e.g.~\cite{IlievskiGGE15,StringCharge}) which are a natural extension of momentum distribution functions 
of free theories~\cite{IQC17}.

\paragraph*{Linearized hydrodynamics.}
The hydrodynamic approach~\cite{BCNF16,CDY16} is based on the notion of local quasi-stationary states, characterized
by the local continuity equation in the mode space $\partial_{t}\rho_{a}(u) + \partial_{x}j_{a}(u) = 0$.
In the simplest scenario, one can think of a quantum quench in which an inhomogeneous initial state is initialized as
two homogeneous equilibrated macroscopic regions brought in contact at $t=0$, see ~\cite{Spohn77,BD14,BD16_review}.
In such a scenario, an emergent nonequilibrium state remains confined to the light cone region determined by particles' dressed
velocities, leading eventually to a quasi-stationary state which depends on the ray coordinate $\zeta = x/t$ and is determined by the 
condition of vanishing convective derivative $(\partial_{t}+v^{\rm dr}_{a}(u)\partial_{x})\vartheta_{a}(u)=0$.
\begin{figure}[t]
\includegraphics[width=0.9\hsize]{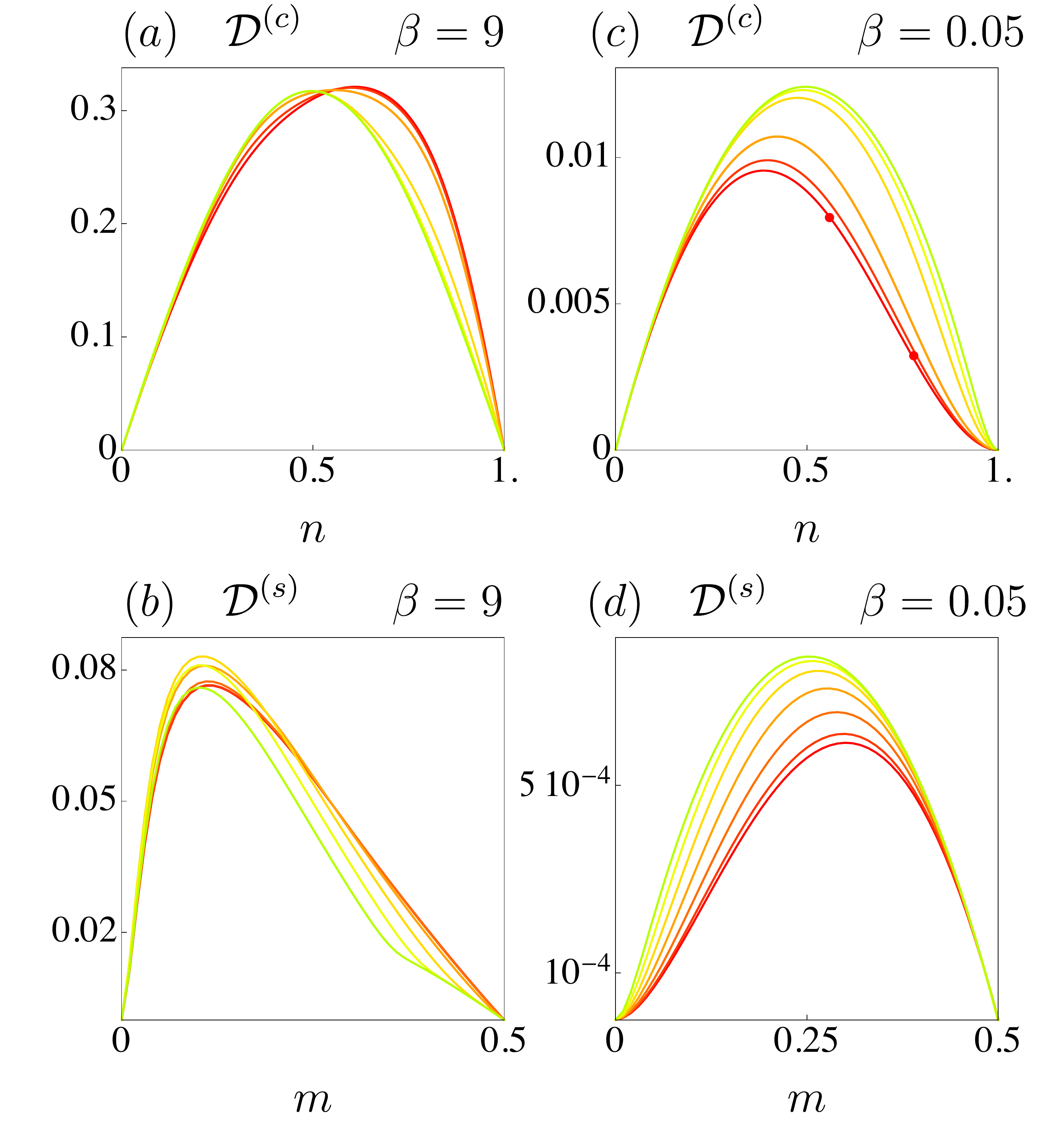}
\caption{Charge Drude $\mathcal{D}^{(c)}\equiv \mathcal{D}^{(c,c)}$ and spin Drude  $\mathcal{D}^{(s)}\equiv \mathcal{D}^{(s,s)}$
weight as functions of magnetization density $\langle \hat{S}^z \rangle/L = m$ or electron filling $\langle \hat{N} \rangle/L = n$, 
shown for different values of chemical potentials: ranging from red to green, with integer $k=0,\ldots,6$, chemical potentials are
parametrized in each plot as (a) $B=2 k$, (b) $\mu = 30 + 5 k$, (c) $B= k$, (d)  $\mu = k$. Red dots are DRMG numerical computations 
reported in \cite{Karrasch17}.}
\label{fig:Charge_D}
\end{figure}

\begin{figure}[b]
\includegraphics[width=0.9\hsize]{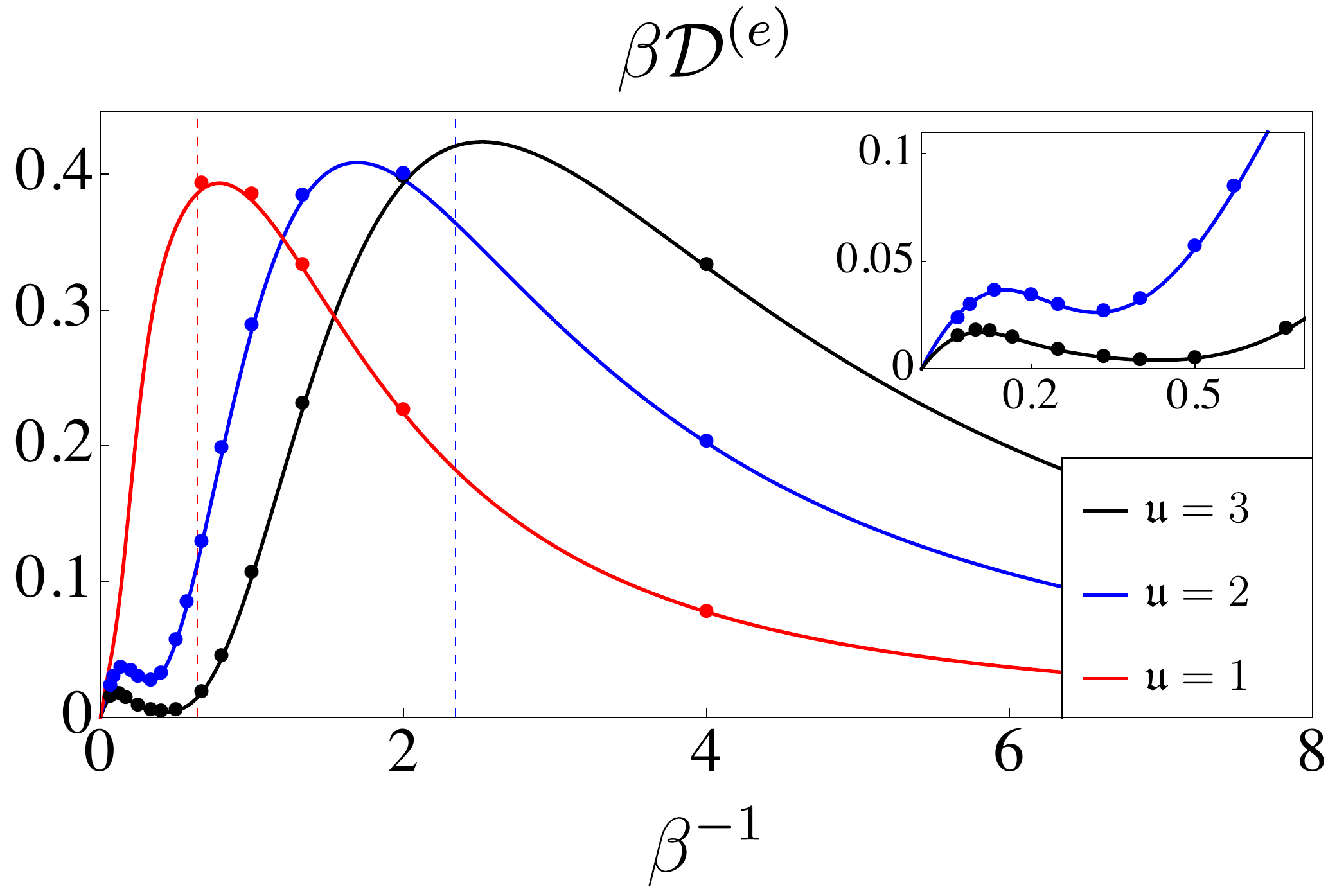}
\caption{Thermal Drude weight $\mathcal{D}^{(e)}\equiv \mathcal{D}^{(e,e)}$ (rescaled by $\beta$) as function of
temperature $1/\beta$, presented for three different values of coupling $\mathfrak{u}$.
The dotted vertical lines represent the charge gap. The inset plot magnifies the region around $1/\beta \sim 0$.
Our results confirm the presence of the low-temperature bump in the thermal Drude weight which comes from the dominant
spin-carrying excitations, suggested and observed numerically in \cite{KKH16,Karrasch17}. The dots drawn on top of the theoretical 
predictions (solid curves) are the results of numerical DMRG calculations presented in \cite{Karrasch17}. }
\label{fig:Therm_D}
\end{figure}

The setting proves particularly useful for studying nonequilibrium transport properties and, in particular, computation
of Drude weights. The latter can be conveniently defined as asymptotic current rates in the limit of vanishing bias $\delta \mu_{j}$
(while keeping other chemical potentials fixed),
\begin{equation}
\calD^{(i,j)} = \frac{\beta}{2}\lim_{\delta \mu_{j}\to 0}\frac{\partial}{\partial\,\delta\mu_{j}}\lim_{t\to \infty}\frac{J_{i}(t)}{t}.
\label{eqn:Drude_from_rates}
\end{equation}
The above prescription has been initially used in \cite{VKM16} and employed in a recent numerical study \cite{Karrasch17}, while
an analogous formula already appeared in an earlier work \cite{IP12}. Equation \eqref{eqn:Drude_from_rates} has been recently 
evaluated in \cite{ID17,BVKM17} using the hydrodynamic approach, transforming it first in the light cone coordinates,
$\calD^{(i,j)} = (\beta/2)\lim_{\delta \mu_{j}\to 0}\int \dd \zeta\,\partial j_{i}(\zeta)/\partial \delta \mu_{j}$, and then
computing quasi-stationary currents which are generated by joining together two nearly identical equilibrium states, i.e.
imposing a small chemical potential drop at the origin $\mu^{\rm L}_{i}=\mu_{i}+\delta \mu_{i}/2$ and
$\mu^{\rm R}_{i}=\mu_{i}-\delta \mu_{i}/2$. Here $\delta \mu_{i}$ has the role of a thermodynamic force, e.g. to study energy
transport we identify $\mu_{e}=\beta$.

Just very recently in \cite{DS17} the authors applied Eq.~\eqref{eqn:Drude_from_rates} to the Lieb--Liniger model and obtained
closed-form expressions analogous to Eq.~\eqref{eqn:resolution}.
Below we generalize this result to interacting quantum models which involve multiple species of 
excitations and internal degrees of freedom. It is quite remarkable however that the final outcome remains a bilinear functional
operating diagonally in the mode-number space, while the effect of interparticle interactions gets absorbed into a universal 
renormalization of bare charges, see Eq.~\eqref{eqn:dressing_abstract}.

Eq.~\eqref{eqn:Drude_from_rates} indicates that Drude weights are expressible as the variation of the equilibrium expectation
values of \emph{total} current \cite{ID17} with respect to thermodynamic forces $\delta \mu_{j}$,
$\calD^{(i,j)} = \tfrac{\beta}{2}(\partial J_{i}/\partial\,\delta \mu_{j})_{\delta \mu_{j}=0} =
\tfrac{\beta}{2}\sum_{a} \iint \dd \zeta\,\dd u\,q_{a,i}(u;\zeta)(\partial j_{a}(u;\zeta)/\partial\,\delta \mu_{j})_{\delta \mu_{j}=0}$, being the susceptibility of a system to develop ballistic currents.
On each ray $\zeta$, the averages of particle current densities are given by \cite{BCNF16}
$j_{a}(\zeta) = (\sigma_{a}\vartheta^{-1}_{a}(\zeta)\delta_{ab}+K_{ab})^{-1}\star \e^{\prime}_{b}(\zeta)$,
where rapidity dependence has been suppressed for brevity. Given the filling functions inside the light cone
$\vartheta_{a}(u;\zeta) = \vartheta^{\rm L}_{a}(u) +
\Theta(v^{\rm dr}_{a}(u)-\zeta)\left(\vartheta^{\rm R}_{a}(u)-\vartheta^{\rm L}_{a}(u)\right)$,
with the left/right boundary conditions $\vartheta^{\rm L,R}_{a}$, and neglecting corrections of order $\mathcal{O}(\delta \mu^{2})$, 
one can integrate out the dependence on the light cone coordinates (see SM~\cite{SM} for details).
This leads to the form of Eq.~\eqref{eqn:resolution}, with
\begin{equation}
D_{a}(u) = \rho_{a}(u)(1-\vartheta_{a}(u))\left(v^{\rm dr}_{a}(u)\right)^{2}.
\label{eqn:Drude_kernel}
\end{equation}

\paragraph*{On detailed balance.}
The symmetry under exchanging indices $i$ an $j$ in representation \eqref{eqn:resolution}, $\calD^{(i,j)}=\calD^{(j,i)}$,
indicates that the Onsager reciprocal relations \cite{Onsager} remain valid for any stationary state,
not only in thermal Gibbs equilibrium. This is indeed a general property of the hydrodynamic equation of motion~\cite{Spohn_book}. Moreover we here show that in a general local equilibrium state
of an integrable quantum model, there exist a generalized \textit{detailed balance} condition on the hydrodynamic
scale (i.e for small $\kappa$ and $\omega$), similarly as in the Lieb-Liniger model found recently 
in~\cite{SciPostPhys.1.2.015,FoiniDeNardis}.
More specifically, given a conserved quantity of the model $\hat{Q}=\sum_x \hat{q}_x$, the corresponding
dynamical structure factor defined as $\calS_{\hat{q}}(\kappa,\omega) = \sum_{x}\int \dd t e^{\ii (\kappa\,x - \omega\,t)}\langle \hat{q}_x(t) \hat{q}_0(0) \rangle$ decomposes in terms of individual particle contributions, $\calS_{\hat{q}}(\kappa,\omega)  = \sum_a \calS_{\hat{q},a}(\kappa,\omega)$. In the low-momentum limit $\kappa \to 0$, each term is determined by a single matrix element
of a particle-hole excitation on a reference equilibrium state~\cite{SM}. Therefore, following the logic presented
in \cite{SciPostPhys.1.2.015}, we derive the following generalized reversibility property 
\begin{equation}
\calS_{\hat{q},a}(\kappa,-\omega)  = e^{-\mathcal{F}_a (k,\omega)}\calS_{\hat{q},a}(\kappa,\omega)  +O(\kappa^2)
\end{equation}
with 
$\mathcal{F}_a(\kappa,\omega)= \kappa \tfrac{\partial}{\partial p_{a}(u)} \log \left( \vartheta_a^{-1} (u) -1 \right) $, with $u$ fixed by the energy constraint $ v_a^{\text{dr}}(u) \kappa = \omega$.  In the case of thermal (canonical) equilibrium, given by
$\vartheta_a = (1 + \exp{(\beta\,\edr_{a} + \sum_{i}\mu_{a,i}n_{a})})^{-1}$, we have $\mathcal{F}_a (\kappa,\omega)=\beta \omega$, 
which is the usual detailed balance relation.

%

\paragraph*{Hubbard model.}

The Hamiltonian of the 1D Hubbard model~\cite{Gutzwiller63,Hubbard63} is given as
\begin{equation}
\hat{H} = \sum_{x=1}^{L}\hat{T}_{x,x+1} +4\uu \sum_{x=1}^{L}\hat{V}_{x,x+1},
\label{eqn:Hubbard}
\end{equation}
where $\hat{T}_{x,x+1} = -\sum_{\sigma=\ua,\da} \hat{c}^{\dagger}_{x,\sigma}\hat{c}_{x+1,\sigma} +
\hat{c}^{\dagger}_{x+1,\sigma}\hat{c}_{x,\sigma}$ is electron hopping and
$\hat{V}_{x,x+1}=(\hat{n}_{x,\ua}-\tfrac{1}{2})(\hat{n}_{x,\da}-\tfrac{1}{2})$ is the Coulomb interaction.
This model has received a lot of attention in the past decades\cite{Shastry88,Woynarovich89,FK90,OS90,EKS91} as well as in the last
years~\cite{PZ12,KKM14,Prosen14,Seabra14,Neumayer15,Reiner16,VE16,Vaness16,Tiegel16,KPH17,Lin17,Karrasch17,Karrasch17HUB,Bolens17}.
We consider the repulsive case $\uu \geq 0$, featuring a $\uu$-dependent charge gap and gapless spin degrees of freedom.

The Hubbard model is diagonalized by means of {nested Bethe Ansatz}~\cite{EKS91,Hubbard_book}.
Eigenstates in a finite system of length $L$ are characterized by quantum numbers which are solutions to
Lieb--Wu equations~\cite{LW68} (cf. SM~\cite{SM})
The model involves two elementary degrees of freedom; the physical particles are momentum-carrying electrons,
while spin degrees of freedom represent internal (non-dynamical) excitations described by auxiliary quantum numbers.
In a thermodynamic system one finds various types of charge and/or spin-carrying bound states.
Specifically, the thermodynamic particle content of the Hubbard model has been derived in \cite{Takahashi_Hubbard} (see
also \cite{Hubbard_book,QF13}) and comprises of (i) spin-up momentum-carrying electronic excitations which carry unit bare 
(electronic) charge (ii) spin-singlet electronic bound states and (iii) charge-neutral \emph{non-dynamical} spin-carrying magnonic 
excitations. A detailed description of the particle content and other information, including explicit expressions for their bare 
momenta, energies, scattering phases and the dressing transformation, are reported in SM~\cite{SM}.

\paragraph*{Numerical results.}
We present temperature dependence of charge and spin, see Fig.~\ref{fig:Charge_D}, and thermal Drude weights, see Fig.~\ref{fig:Therm_D}, in grand canonical equilibrium
$\hat{\varrho}_{GCE}\simeq \exp{(-\beta\,\hat{H}-\mu\,\hat{N}+B\,\hat{S}^{\rm z})}$, where
$\hat{N} = \sum_{x=1}^{L}(\hat{c}^{\dagger}_{x,\ua}\hat{c}_{x,\ua} + \hat{c}^{\dagger}_{x,\da}\hat{c}_{x,\da})$ is
total electron charge, and
$\hat{S}^{z} = \tfrac{1}{2}\sum_{x=1}^{L}(\hat{c}^{\dagger}_{x,\ua}\hat{c}_{i,\ua}-\hat{c}^{\dagger}_{x,\da}\hat{c}_{x,\da})$,
total magnetization.
We compared our data with the recent DMRG computation presented in \cite{Karrasch17HUB,Karrasch17}. Most notably, 
at low temperatures appreciably below the charge gap we confirm the `Hubbard to Heisenberg crossover' in the thermal Drude weight 
observed previously in \cite{KKH16,Karrasch17}, see Fig.~\ref{fig:Therm_D}. In \cite{SM} we also present an exact computation of the 
asymptotic charge and current profiles inside a light cone and make comparisons with the numerical results of~\cite{Karrasch17HUB}.

\paragraph*{Conclusions.}
We presented a general theoretical and computational framework to access the singular components (Drude weights) of generalized 
transport coefficients in quantum integrable models. We exemplified our approach by computing exact numerical
values of (diagonal) charge, spin and thermal Drude weights in the one-dimensional fermionic Hubbard model in grand canonical 
equilibrium at finite temperatures and chemical potentials. Using the two-partition protocol, we additionally
computed the quasi-stationary energy and charge density profile and the corresponding current ~\cite{SM}.

Our results finally permit to establish the equivalence of various approaches for computing the spin Drude 
weight employed in the previous literature: (i) using projections onto local conserved subspaces by virtue of 
Mazur--Suzuki equality~\cite{ZNP97,Prosen11,PI13}, (ii) taking the linear-response limit of the asymptotic current 
rates~\cite{ID17,BVKM17} and (iii) computing the energy-level curvatures~\cite{SS90,FK98,Zotos99,BFKS05} under the twisted boundary 
conditions in accordance with Kohn formula~\cite{Kohn64}. The latter has been evaluated within the TBA framework
in \cite{FK98,Zotos99}, yielding a closed formula expressed in terms of filling functions, magnonic dispersion relations
and $\mathcal{O}(1/L)$ corrections to the Bethe spectrum induced by the twist.
Remarkably however, it is easy to see that the twist-dependence of the energy levels can be directly linked
to the effective spin as given by Eq.~\eqref{eqn:dressing_transformation}. This in turn reconciles the results of \cite{Zotos99} with 
Eq.~\eqref{eqn:Drude_kernel}, representing the equilibrium analogue of definition \eqref{eqn:Drude_from_rates} used previously
in refs.~\cite{ID17,BVKM17} (further details are given in SM~\cite{SM}, which also includes 
refs.~\cite{Narozhny98,Peres99,AG02,HM03,FK03,Herbrych11,SPA11,Znidaric11,Gromov11,KBM12,Cavaglia15,DeLuca16}).

Finally, our results show that a generalized version of the detailed
balance \cite{SciPostPhys.1.2.015,FoiniDeNardis,PhysRevE.95.052116} is valid on hydrodynamic scales in any stationary state.

As a future task, it would be interesting to find an extension of the presented approach which would allow
resolving the diffusive time-scale from the microscopic picture, see e.g.~\cite{MKP17,Medenjak17}.

\paragraph*{Authors contributions.}
Both authors contributed equally to the theory. J. De Nardis performed the numerical computations.

\paragraph*{Acknowledgements.}
We are grateful to C. Karrasch for providing the tDMRG data for the Drude weights in Hubbard model
and thank M. Van Caspel, M. Fagotti, E. Quinn and H. Spohn for valuable discussions and/or reading the manuscript.
E.I. is supported by VENI grant number 680-47-454 by the Netherlands Organisation for Scientific Research (NWO).
J.D.N. acknowledge support by LabEx ENS-ICFP:ANR-10-LABX-0010/ANR-10-IDEX-0001-02 PSL*.
\bibliography{Hubbard}

\begin{thebibliography}{97}%
\makeatletter
\providecommand \@ifxundefined [1]{%
 \@ifx{#1\undefined}
}%
\providecommand \@ifnum [1]{%
 \ifnum #1\expandafter \@firstoftwo
 \else \expandafter \@secondoftwo
 \fi
}%
\providecommand \@ifx [1]{%
 \ifx #1\expandafter \@firstoftwo
 \else \expandafter \@secondoftwo
 \fi
}%
\providecommand \natexlab [1]{#1}%
\providecommand \enquote  [1]{``#1''}%
\providecommand \bibnamefont  [1]{#1}%
\providecommand \bibfnamefont [1]{#1}%
\providecommand \citenamefont [1]{#1}%
\providecommand \href@noop [0]{\@secondoftwo}%
\providecommand \href [0]{\begingroup \@sanitize@url \@href}%
\providecommand \@href[1]{\@@startlink{#1}\@@href}%
\providecommand \@@href[1]{\endgroup#1\@@endlink}%
\providecommand \@sanitize@url [0]{\catcode `\\12\catcode `\$12\catcode
  `\&12\catcode `\#12\catcode `\^12\catcode `\_12\catcode `\%12\relax}%
\providecommand \@@startlink[1]{}%
\providecommand \@@endlink[0]{}%
\providecommand \url  [0]{\begingroup\@sanitize@url \@url }%
\providecommand \@url [1]{\endgroup\@href {#1}{\urlprefix }}%
\providecommand \urlprefix  [0]{URL }%
\providecommand \Eprint [0]{\href }%
\providecommand \doibase [0]{http://dx.doi.org/}%
\providecommand \selectlanguage [0]{\@gobble}%
\providecommand \bibinfo  [0]{\@secondoftwo}%
\providecommand \bibfield  [0]{\@secondoftwo}%
\providecommand \translation [1]{[#1]}%
\providecommand \BibitemOpen [0]{}%
\providecommand \bibitemStop [0]{}%
\providecommand \bibitemNoStop [0]{.\EOS\space}%
\providecommand \EOS [0]{\spacefactor3000\relax}%
\providecommand \BibitemShut  [1]{\csname bibitem#1\endcsname}%
\let\auto@bib@innerbib\@empty
\bibitem [{\citenamefont {Polkovnikov}\ \emph {et~al.}(2011)\citenamefont
  {Polkovnikov}, \citenamefont {Sengupta}, \citenamefont {Silva},\ and\
  \citenamefont {Vengalattore}}]{Polkovnikov_review}%
  \BibitemOpen
  \bibfield  {author} {\bibinfo {author} {\bibfnamefont {A.}~\bibnamefont
  {Polkovnikov}}, \bibinfo {author} {\bibfnamefont {K.}~\bibnamefont
  {Sengupta}}, \bibinfo {author} {\bibfnamefont {A.}~\bibnamefont {Silva}}, \
  and\ \bibinfo {author} {\bibfnamefont {M.}~\bibnamefont {Vengalattore}},\
  }\href {\doibase 10.1103/revmodphys.83.863} {\bibfield  {journal} {\bibinfo
  {journal} {Reviews of Modern Physics}\ }\textbf {\bibinfo {volume} {83}},\
  \bibinfo {pages} {863} (\bibinfo {year} {2011})}\BibitemShut {NoStop}%
\bibitem [{\citenamefont {Calabrese}\ \emph {et~al.}(2016)\citenamefont
  {Calabrese}, \citenamefont {Essler},\ and\ \citenamefont
  {Mussardo}}]{Calabrese_intro}%
  \BibitemOpen
  \bibfield  {author} {\bibinfo {author} {\bibfnamefont {P.}~\bibnamefont
  {Calabrese}}, \bibinfo {author} {\bibfnamefont {F.~H.~L.}\ \bibnamefont
  {Essler}}, \ and\ \bibinfo {author} {\bibfnamefont {G.}~\bibnamefont
  {Mussardo}},\ }\href {\doibase 10.1088/1742-5468/2016/06/064001} {\bibfield
  {journal} {\bibinfo  {journal} {Journal of Statistical Mechanics: Theory and
  Experiment}\ }\textbf {\bibinfo {volume} {2016}},\ \bibinfo {pages} {064001}
  (\bibinfo {year} {2016})}\BibitemShut {NoStop}%
\bibitem [{\citenamefont {Gogolin}\ and\ \citenamefont
  {Eisert}(2016)}]{Gogolin}%
  \BibitemOpen
  \bibfield  {author} {\bibinfo {author} {\bibfnamefont {C.}~\bibnamefont
  {Gogolin}}\ and\ \bibinfo {author} {\bibfnamefont {J.}~\bibnamefont
  {Eisert}},\ }\href {\doibase 10.1088/0034-4885/79/5/056001} {\bibfield
  {journal} {\bibinfo  {journal} {Reports on Progress in Physics}\ }\textbf
  {\bibinfo {volume} {79}},\ \bibinfo {pages} {056001} (\bibinfo {year}
  {2016})}\BibitemShut {NoStop}%
\bibitem [{\citenamefont {Bertini}\ \emph {et~al.}(2016)\citenamefont
  {Bertini}, \citenamefont {Collura}, \citenamefont {Nardis},\ and\
  \citenamefont {Fagotti}}]{BCNF16}%
  \BibitemOpen
  \bibfield  {author} {\bibinfo {author} {\bibfnamefont {B.}~\bibnamefont
  {Bertini}}, \bibinfo {author} {\bibfnamefont {M.}~\bibnamefont {Collura}},
  \bibinfo {author} {\bibfnamefont {J.~D.}\ \bibnamefont {Nardis}}, \ and\
  \bibinfo {author} {\bibfnamefont {M.}~\bibnamefont {Fagotti}},\ }\href
  {\doibase 10.1103/physrevlett.117.207201} {\bibfield  {journal} {\bibinfo
  {journal} {Physical Review Letters}\ }\textbf {\bibinfo {volume} {117}}
  (\bibinfo {year} {2016}),\ 10.1103/physrevlett.117.207201}\BibitemShut
  {NoStop}%
\bibitem [{\citenamefont {Castro-Alvaredo}\ \emph {et~al.}(2016)\citenamefont
  {Castro-Alvaredo}, \citenamefont {Doyon},\ and\ \citenamefont
  {Yoshimura}}]{CDY16}%
  \BibitemOpen
  \bibfield  {author} {\bibinfo {author} {\bibfnamefont {O.~A.}\ \bibnamefont
  {Castro-Alvaredo}}, \bibinfo {author} {\bibfnamefont {B.}~\bibnamefont
  {Doyon}}, \ and\ \bibinfo {author} {\bibfnamefont {T.}~\bibnamefont
  {Yoshimura}},\ }\href {\doibase 10.1103/physrevx.6.041065} {\bibfield
  {journal} {\bibinfo  {journal} {Physical Review X}\ }\textbf {\bibinfo
  {volume} {6}} (\bibinfo {year} {2016}),\
  10.1103/physrevx.6.041065}\BibitemShut {NoStop}%
\bibitem [{\citenamefont {Doyon}\ and\ \citenamefont
  {Yoshimura}(2017)}]{DoYo16}%
  \BibitemOpen
  \bibfield  {author} {\bibinfo {author} {\bibfnamefont {B.}~\bibnamefont
  {Doyon}}\ and\ \bibinfo {author} {\bibfnamefont {T.}~\bibnamefont
  {Yoshimura}},\ }\href {\doibase 10.21468/SciPostPhys.2.2.014} {\bibfield
  {journal} {\bibinfo  {journal} {SciPost Phys.}\ }\textbf {\bibinfo {volume}
  {2}},\ \bibinfo {pages} {014} (\bibinfo {year} {2017})}\BibitemShut {NoStop}%
\bibitem [{\citenamefont {{B. Doyon, and H. Spohn, and T.
  Yoshimura}}(2017)}]{DoSY17}%
  \BibitemOpen
  \bibfield  {author} {\bibinfo {author} {\bibnamefont {{B. Doyon, and H.
  Spohn, and T. Yoshimura}}},\ }\href {http://arxiv.org/abs/1704.04409}
  {\bibfield  {journal} {\bibinfo  {journal} {arXiv preprint arXiv:1704.04409}\
  } (\bibinfo {year} {2017})}\BibitemShut {NoStop}%
\bibitem [{\citenamefont {{B. Doyon and H. Spohn}}(2017)}]{DoSp17}%
  \BibitemOpen
  \bibfield  {author} {\bibinfo {author} {\bibnamefont {{B. Doyon and H.
  Spohn}}},\ }\href {http://arxiv.org/abs/1703.05971} {\bibfield  {journal}
  {\bibinfo  {journal} {arXiv preprint arXiv:1703.05971}\ } (\bibinfo {year}
  {2017})}\BibitemShut {NoStop}%
\bibitem [{\citenamefont {{B. Doyon, T. Yoshimura, and J.-S.
  Caux}}(2017)}]{DoYC17}%
  \BibitemOpen
  \bibfield  {author} {\bibinfo {author} {\bibnamefont {{B. Doyon, T.
  Yoshimura, and J.-S. Caux}}},\ }\href {http://arxiv.org/abs/1704.05482}
  {\bibfield  {journal} {\bibinfo  {journal} {arXiv preprint arXiv:1704.05482}\
  } (\bibinfo {year} {2017})}\BibitemShut {NoStop}%
\bibitem [{\citenamefont {{B. Doyon, J. Dubail, R. Konik, and T.
  Yoshimura}}(2017)}]{DDKY17}%
  \BibitemOpen
  \bibfield  {author} {\bibinfo {author} {\bibnamefont {{B. Doyon, J. Dubail,
  R. Konik, and T. Yoshimura}}},\ }\href {http://arxiv.org/abs/1704.04151}
  {\bibfield  {journal} {\bibinfo  {journal} {arXiv preprint arXiv:1704.04151}\
  } (\bibinfo {year} {2017})}\BibitemShut {NoStop}%
\bibitem [{\citenamefont {{V. B. Bulchandani, R. Vasseur, C. Karrasch, and J.
  E. Moore}}(2017)}]{BVKM17-2}%
  \BibitemOpen
  \bibfield  {author} {\bibinfo {author} {\bibnamefont {{V. B. Bulchandani, R.
  Vasseur, C. Karrasch, and J. E. Moore}}},\ }\href
  {http://arxiv.org/abs/1704.03466} {\bibfield  {journal} {\bibinfo  {journal}
  {arXiv preprint arXiv:1704.03466}\ } (\bibinfo {year} {2017})}\BibitemShut
  {NoStop}%
\bibitem [{\citenamefont {{Piroli, Lorenzo and Nardis, Jacopo De and Collura,
  Mario and Bertini, Bruno and Fagotti, Maurizio}}(2017)}]{Piroli-longBertini}%
  \BibitemOpen
  \bibfield  {author} {\bibinfo {author} {\bibnamefont {{Piroli, Lorenzo and
  Nardis, Jacopo De and Collura, Mario and Bertini, Bruno and Fagotti,
  Maurizio}}},\ }\href {https://arxiv.org/abs/1706.00413} {\bibfield  {journal}
  {\bibinfo  {journal} {arXiv preprint arXiv:1706.00413}\ } (\bibinfo {year}
  {2017})}\BibitemShut {NoStop}%
\bibitem [{\citenamefont {{Alba, Vincenzo}}(2017)}]{VAlba_EntropyTransport}%
  \BibitemOpen
  \bibfield  {author} {\bibinfo {author} {\bibnamefont {{Alba, Vincenzo}}},\
  }\href {https://arxiv.org/abs/1706.00020} {\bibfield  {journal} {\bibinfo
  {journal} {arXiv preprint arXiv:1706.00020}\ } (\bibinfo {year}
  {2017})}\BibitemShut {NoStop}%
\bibitem [{\citenamefont {Ilievski}\ and\ \citenamefont
  {De~Nardis}(2017)}]{ID17}%
  \BibitemOpen
  \bibfield  {author} {\bibinfo {author} {\bibfnamefont {E.}~\bibnamefont
  {Ilievski}}\ and\ \bibinfo {author} {\bibfnamefont {J.}~\bibnamefont
  {De~Nardis}},\ }\href {https://arxiv.org/abs/1702.02930} {\bibfield
  {journal} {\bibinfo  {journal} {arXiv preprint arXiv:1702.02930}\ } (\bibinfo
  {year} {2017})}\BibitemShut {NoStop}%
\bibitem [{\citenamefont {Spohn}(1991)}]{Spohn_book}%
  \BibitemOpen
  \bibfield  {author} {\bibinfo {author} {\bibfnamefont {H.}~\bibnamefont
  {Spohn}},\ }\href {\doibase 10.1007/978-3-642-84371-6} {\emph {\bibinfo
  {title} {Large Scale Dynamics of Interacting Particles}}}\ (\bibinfo
  {publisher} {Springer Berlin Heidelberg},\ \bibinfo {year}
  {1991})\BibitemShut {NoStop}%
\bibitem [{\citenamefont {Doyon}\ and\ \citenamefont {Spohn}(2017)}]{DS17}%
  \BibitemOpen
  \bibfield  {author} {\bibinfo {author} {\bibfnamefont {B.}~\bibnamefont
  {Doyon}}\ and\ \bibinfo {author} {\bibfnamefont {H.}~\bibnamefont {Spohn}},\
  }\href@noop {} {\bibfield  {journal} {\bibinfo  {journal} {arXiv preprint
  arXiv:1705.08141}\ } (\bibinfo {year} {2017})}\BibitemShut {NoStop}%
\bibitem [{\citenamefont {Essler}\ \emph {et~al.}(1991)\citenamefont {Essler},
  \citenamefont {Korepin},\ and\ \citenamefont {Schoutens}}]{EKS91}%
  \BibitemOpen
  \bibfield  {author} {\bibinfo {author} {\bibfnamefont {F.~H.~L.}\
  \bibnamefont {Essler}}, \bibinfo {author} {\bibfnamefont {V.~E.}\
  \bibnamefont {Korepin}}, \ and\ \bibinfo {author} {\bibfnamefont
  {K.}~\bibnamefont {Schoutens}},\ }\href {\doibase
  10.1103/physrevlett.67.3848} {\bibfield  {journal} {\bibinfo  {journal}
  {Physical Review Letters}\ }\textbf {\bibinfo {volume} {67}},\ \bibinfo
  {pages} {3848} (\bibinfo {year} {1991})}\BibitemShut {NoStop}%
\bibitem [{\citenamefont {Essler}\ \emph {et~al.}(1992)\citenamefont {Essler},
  \citenamefont {Korepin},\ and\ \citenamefont {Schoutens}}]{EKS92}%
  \BibitemOpen
  \bibfield  {author} {\bibinfo {author} {\bibfnamefont {F.~H.~L.}\
  \bibnamefont {Essler}}, \bibinfo {author} {\bibfnamefont {V.~E.}\
  \bibnamefont {Korepin}}, \ and\ \bibinfo {author} {\bibfnamefont
  {K.}~\bibnamefont {Schoutens}},\ }\href {\doibase
  10.1103/physrevlett.68.2960} {\bibfield  {journal} {\bibinfo  {journal}
  {Physical Review Letters}\ }\textbf {\bibinfo {volume} {68}},\ \bibinfo
  {pages} {2960} (\bibinfo {year} {1992})}\BibitemShut {NoStop}%
\bibitem [{\citenamefont {Essler}\ and\ \citenamefont {Korepin}(1994)}]{EK94}%
  \BibitemOpen
  \bibfield  {author} {\bibinfo {author} {\bibfnamefont {F.~H.}\ \bibnamefont
  {Essler}}\ and\ \bibinfo {author} {\bibfnamefont {V.~E.}\ \bibnamefont
  {Korepin}},\ }\href {\doibase 10.1142/s0217979294001366} {\bibfield
  {journal} {\bibinfo  {journal} {International Journal of Modern Physics B}\
  }\textbf {\bibinfo {volume} {08}},\ \bibinfo {pages} {3243} (\bibinfo {year}
  {1994})}\BibitemShut {NoStop}%
\bibitem [{\citenamefont {{Fabian H. L. Essler and Holger Frahm and Frank
  G\"{o}hmann and Andreas Kl\"{u}mper and Vladimir E.
  Korepin}}(2005)}]{Hubbard_book}%
  \BibitemOpen
  \bibfield  {author} {\bibinfo {author} {\bibnamefont {{Fabian H. L. Essler
  and Holger Frahm and Frank G\"{o}hmann and Andreas Kl\"{u}mper and Vladimir
  E. Korepin}}},\ }\href {\doibase 10.1017/cbo9780511534843} {\emph {\bibinfo
  {title} {{The One-Dimensional Hubbard Model}}}}\ (\bibinfo  {publisher}
  {Cambridge University Press ({CUP})},\ \bibinfo {year} {2005})\BibitemShut
  {NoStop}%
\bibitem [{\citenamefont {Beisert}\ \emph {et~al.}(2011)\citenamefont
  {Beisert}, \citenamefont {Ahn}, \citenamefont {Alday}, \citenamefont
  {Bajnok}, \citenamefont {Drummond}, \citenamefont {Freyhult}, \citenamefont
  {Gromov}, \citenamefont {Janik}, \citenamefont {Kazakov}, \citenamefont
  {Klose}, \citenamefont {Korchemsky}, \citenamefont {Kristjansen},
  \citenamefont {Magro}, \citenamefont {McLoughlin}, \citenamefont {Minahan},
  \citenamefont {Nepomechie}, \citenamefont {Rej}, \citenamefont {Roiban},
  \citenamefont {Schäfer-Nameki}, \citenamefont {Sieg}, \citenamefont
  {Staudacher}, \citenamefont {Torrielli}, \citenamefont {Tseytlin},
  \citenamefont {Vieira}, \citenamefont {Volin},\ and\ \citenamefont
  {Zoubos}}]{AdSCFT_review}%
  \BibitemOpen
  \bibfield  {author} {\bibinfo {author} {\bibfnamefont {N.}~\bibnamefont
  {Beisert}}, \bibinfo {author} {\bibfnamefont {C.}~\bibnamefont {Ahn}},
  \bibinfo {author} {\bibfnamefont {L.~F.}\ \bibnamefont {Alday}}, \bibinfo
  {author} {\bibfnamefont {Z.}~\bibnamefont {Bajnok}}, \bibinfo {author}
  {\bibfnamefont {J.~M.}\ \bibnamefont {Drummond}}, \bibinfo {author}
  {\bibfnamefont {L.}~\bibnamefont {Freyhult}}, \bibinfo {author}
  {\bibfnamefont {N.}~\bibnamefont {Gromov}}, \bibinfo {author} {\bibfnamefont
  {R.~A.}\ \bibnamefont {Janik}}, \bibinfo {author} {\bibfnamefont
  {V.}~\bibnamefont {Kazakov}}, \bibinfo {author} {\bibfnamefont
  {T.}~\bibnamefont {Klose}}, \bibinfo {author} {\bibfnamefont {G.~P.}\
  \bibnamefont {Korchemsky}}, \bibinfo {author} {\bibfnamefont
  {C.}~\bibnamefont {Kristjansen}}, \bibinfo {author} {\bibfnamefont
  {M.}~\bibnamefont {Magro}}, \bibinfo {author} {\bibfnamefont
  {T.}~\bibnamefont {McLoughlin}}, \bibinfo {author} {\bibfnamefont {J.~A.}\
  \bibnamefont {Minahan}}, \bibinfo {author} {\bibfnamefont {R.~I.}\
  \bibnamefont {Nepomechie}}, \bibinfo {author} {\bibfnamefont
  {A.}~\bibnamefont {Rej}}, \bibinfo {author} {\bibfnamefont {R.}~\bibnamefont
  {Roiban}}, \bibinfo {author} {\bibfnamefont {S.}~\bibnamefont
  {Schäfer-Nameki}}, \bibinfo {author} {\bibfnamefont {C.}~\bibnamefont
  {Sieg}}, \bibinfo {author} {\bibfnamefont {M.}~\bibnamefont {Staudacher}},
  \bibinfo {author} {\bibfnamefont {A.}~\bibnamefont {Torrielli}}, \bibinfo
  {author} {\bibfnamefont {A.~A.}\ \bibnamefont {Tseytlin}}, \bibinfo {author}
  {\bibfnamefont {P.}~\bibnamefont {Vieira}}, \bibinfo {author} {\bibfnamefont
  {D.}~\bibnamefont {Volin}}, \ and\ \bibinfo {author} {\bibfnamefont
  {K.}~\bibnamefont {Zoubos}},\ }\href {\doibase 10.1007/s11005-011-0529-2}
  {\bibfield  {journal} {\bibinfo  {journal} {Letters in Mathematical Physics}\
  }\textbf {\bibinfo {volume} {99}},\ \bibinfo {pages} {3} (\bibinfo {year}
  {2011})}\BibitemShut {NoStop}%
\bibitem [{\citenamefont {Frolov}\ and\ \citenamefont {Quinn}(2012)}]{FQ12}%
  \BibitemOpen
  \bibfield  {author} {\bibinfo {author} {\bibfnamefont {S.}~\bibnamefont
  {Frolov}}\ and\ \bibinfo {author} {\bibfnamefont {E.}~\bibnamefont {Quinn}},\
  }\href {\doibase 10.1088/1751-8113/45/9/095004} {\bibfield  {journal}
  {\bibinfo  {journal} {Journal of Physics A: Mathematical and Theoretical}\
  }\textbf {\bibinfo {volume} {45}},\ \bibinfo {pages} {095004} (\bibinfo
  {year} {2012})}\BibitemShut {NoStop}%
\bibitem [{\citenamefont {{Mestyan, Marton and Bertini, Bruno and Piroli,
  Lorenzo and Calabrese, Pasquale}}(2017)}]{CalabreseNested}%
  \BibitemOpen
  \bibfield  {author} {\bibinfo {author} {\bibnamefont {{Mestyan, Marton and
  Bertini, Bruno and Piroli, Lorenzo and Calabrese, Pasquale}}},\ }\href
  {https://arxiv.org/abs/1705.00851} {\bibfield  {journal} {\bibinfo  {journal}
  {arXiv preprint arXiv:1705.00851}\ } (\bibinfo {year} {2017})}\BibitemShut
  {NoStop}%
\bibitem [{\citenamefont {{Robinson, Neil J. and Caux, Jean-S\'ebastien and
  Konik, M. Robert }}(2016)}]{RCK16}%
  \BibitemOpen
  \bibfield  {author} {\bibinfo {author} {\bibnamefont {{Robinson, Neil J. and
  Caux, Jean-S\'ebastien and Konik, M. Robert }}},\ }\href
  {https://arxiv.org/abs/1602.05532} {\bibfield  {journal} {\bibinfo  {journal}
  {arXiv preprint arXiv:1602.05532}\ } (\bibinfo {year} {2016})}\BibitemShut
  {NoStop}%
\bibitem [{\citenamefont {Korepin}\ \emph {et~al.}(1993)\citenamefont
  {Korepin}, \citenamefont {Bogoliubov},\ and\ \citenamefont
  {Izergin}}]{Korepin_book}%
  \BibitemOpen
  \bibfield  {author} {\bibinfo {author} {\bibfnamefont {V.~E.}\ \bibnamefont
  {Korepin}}, \bibinfo {author} {\bibfnamefont {N.~M.}\ \bibnamefont
  {Bogoliubov}}, \ and\ \bibinfo {author} {\bibfnamefont {A.~G.}\ \bibnamefont
  {Izergin}},\ }\href {\doibase 10.1017/cbo9780511628832} {\emph {\bibinfo
  {title} {{Quantum Inverse Scattering Method and Correlation Functions}}}}\
  (\bibinfo  {publisher} {Cambridge University Press},\ \bibinfo {year}
  {1993})\BibitemShut {NoStop}%
\bibitem [{\citenamefont {Bares}\ \emph {et~al.}(1991)\citenamefont {Bares},
  \citenamefont {Blatter},\ and\ \citenamefont {Ogata}}]{Bares91}%
  \BibitemOpen
  \bibfield  {author} {\bibinfo {author} {\bibfnamefont {P.-A.}\ \bibnamefont
  {Bares}}, \bibinfo {author} {\bibfnamefont {G.}~\bibnamefont {Blatter}}, \
  and\ \bibinfo {author} {\bibfnamefont {M.}~\bibnamefont {Ogata}},\ }\href
  {\doibase 10.1103/physrevb.44.130} {\bibfield  {journal} {\bibinfo  {journal}
  {Physical Review B}\ }\textbf {\bibinfo {volume} {44}},\ \bibinfo {pages}
  {130} (\bibinfo {year} {1991})}\BibitemShut {NoStop}%
\bibitem [{\citenamefont {Quinn}\ and\ \citenamefont {Frolov}(2013)}]{QF13}%
  \BibitemOpen
  \bibfield  {author} {\bibinfo {author} {\bibfnamefont {E.}~\bibnamefont
  {Quinn}}\ and\ \bibinfo {author} {\bibfnamefont {S.}~\bibnamefont {Frolov}},\
  }\href {\doibase 10.1088/1751-8113/46/20/205001} {\bibfield  {journal}
  {\bibinfo  {journal} {Journal of Physics A: Mathematical and Theoretical}\
  }\textbf {\bibinfo {volume} {46}},\ \bibinfo {pages} {205001} (\bibinfo
  {year} {2013})}\BibitemShut {NoStop}%
\bibitem [{\citenamefont {Zamolodchikov}\ and\ \citenamefont
  {Zamolodchikov}(1979)}]{ZZ79}%
  \BibitemOpen
  \bibfield  {author} {\bibinfo {author} {\bibfnamefont {A.~B.}\ \bibnamefont
  {Zamolodchikov}}\ and\ \bibinfo {author} {\bibfnamefont {A.~B.}\ \bibnamefont
  {Zamolodchikov}},\ }\href {\doibase 10.1016/0003-4916(79)90261-6} {\bibfield
  {journal} {\bibinfo  {journal} {Annals of physics}\ }\textbf {\bibinfo
  {volume} {120}},\ \bibinfo {pages} {253} (\bibinfo {year}
  {1979})}\BibitemShut {NoStop}%
\bibitem [{Note1()}]{Note1}%
  \BibitemOpen
  \bibinfo {note} {In fermionic interacting integrable models there exist
  distinct inequivalent possibilities of choosing a bare vacuum. Despite this
  results in different sets of excitations, various choices have no effect on
  the values of physical observables.}\BibitemShut {Stop}%
\bibitem [{\citenamefont {Vidmar}\ and\ \citenamefont
  {Rigol}(2016)}]{VR16_review}%
  \BibitemOpen
  \bibfield  {author} {\bibinfo {author} {\bibfnamefont {L.}~\bibnamefont
  {Vidmar}}\ and\ \bibinfo {author} {\bibfnamefont {M.}~\bibnamefont {Rigol}},\
  }\href {\doibase 10.1088/1742-5468/2016/06/064007} {\bibfield  {journal}
  {\bibinfo  {journal} {J. Stat. Mech.}\ }\textbf {\bibinfo {volume} {2016}},\
  \bibinfo {pages} {064007} (\bibinfo {year} {2016})}\BibitemShut {NoStop}%
\bibitem [{\citenamefont {Essler}\ and\ \citenamefont
  {Fagotti}(2016)}]{EF16_review}%
  \BibitemOpen
  \bibfield  {author} {\bibinfo {author} {\bibfnamefont {F.~H.~L.}\
  \bibnamefont {Essler}}\ and\ \bibinfo {author} {\bibfnamefont
  {M.}~\bibnamefont {Fagotti}},\ }\href {\doibase
  10.1088/1742-5468/2016/06/064002} {\bibfield  {journal} {\bibinfo  {journal}
  {J. Stat. Mech.}\ }\textbf {\bibinfo {volume} {2016}},\ \bibinfo {pages}
  {064002} (\bibinfo {year} {2016})}\BibitemShut {NoStop}%
\bibitem [{\citenamefont {Ilievski}\ \emph {et~al.}(2015)\citenamefont
  {Ilievski}, \citenamefont {De~Nardis}, \citenamefont {Wouters}, \citenamefont
  {Caux}, \citenamefont {Essler},\ and\ \citenamefont
  {Prosen}}]{IlievskiGGE15}%
  \BibitemOpen
  \bibfield  {author} {\bibinfo {author} {\bibfnamefont {E.}~\bibnamefont
  {Ilievski}}, \bibinfo {author} {\bibfnamefont {J.}~\bibnamefont {De~Nardis}},
  \bibinfo {author} {\bibfnamefont {B.}~\bibnamefont {Wouters}}, \bibinfo
  {author} {\bibfnamefont {J.-S.}\ \bibnamefont {Caux}}, \bibinfo {author}
  {\bibfnamefont {F.~H.}\ \bibnamefont {Essler}}, \ and\ \bibinfo {author}
  {\bibfnamefont {T.}~\bibnamefont {Prosen}},\ }\href {\doibase
  10.1103/physrevlett.115.157201} {\bibfield  {journal} {\bibinfo  {journal}
  {Physical Review Letters}\ }\textbf {\bibinfo {volume} {115}},\ \bibinfo
  {pages} {157201} (\bibinfo {year} {2015})}\BibitemShut {NoStop}%
\bibitem [{SM()}]{SM}%
  \BibitemOpen
  \href@noop {} {}\bibinfo {note} {Supplemental Material associated with this
  manuscript.}\BibitemShut {Stop}%
\bibitem [{\citenamefont {Kubo}(1957)}]{Kubo57}%
  \BibitemOpen
  \bibfield  {author} {\bibinfo {author} {\bibfnamefont {R.}~\bibnamefont
  {Kubo}},\ }\href {\doibase 10.1143/jpsj.12.570} {\bibfield  {journal}
  {\bibinfo  {journal} {Journal of the Physical Society of Japan}\ }\textbf
  {\bibinfo {volume} {12}},\ \bibinfo {pages} {570} (\bibinfo {year}
  {1957})}\BibitemShut {NoStop}%
\bibitem [{\citenamefont {Mahan}(2000)}]{Mahan_book}%
  \BibitemOpen
  \bibfield  {author} {\bibinfo {author} {\bibfnamefont {G.~D.}\ \bibnamefont
  {Mahan}},\ }\href {\doibase 10.1007/978-1-4757-5714-9} {\emph {\bibinfo
  {title} {{Many-Particle Physics}}}}\ (\bibinfo  {publisher} {Springer
  Nature},\ \bibinfo {year} {2000})\BibitemShut {NoStop}%
\bibitem [{\citenamefont {Mazur}(1969)}]{Mazur69}%
  \BibitemOpen
  \bibfield  {author} {\bibinfo {author} {\bibfnamefont {P.}~\bibnamefont
  {Mazur}},\ }\href {\doibase 10.1016/0031-8914(69)90185-2} {\bibfield
  {journal} {\bibinfo  {journal} {Physica}\ }\textbf {\bibinfo {volume} {43}},\
  \bibinfo {pages} {533} (\bibinfo {year} {1969})}\BibitemShut {NoStop}%
\bibitem [{\citenamefont {Suzuki}(1971)}]{Suzuki71}%
  \BibitemOpen
  \bibfield  {author} {\bibinfo {author} {\bibfnamefont {M.}~\bibnamefont
  {Suzuki}},\ }\href {\doibase 10.1016/0031-8914(71)90226-6} {\bibfield
  {journal} {\bibinfo  {journal} {Physica}\ }\textbf {\bibinfo {volume} {51}},\
  \bibinfo {pages} {277} (\bibinfo {year} {1971})}\BibitemShut {NoStop}%
\bibitem [{\citenamefont {Zotos}\ \emph {et~al.}(1997)\citenamefont {Zotos},
  \citenamefont {Naef},\ and\ \citenamefont {Prelov\v{s}ek}}]{ZNP97}%
  \BibitemOpen
  \bibfield  {author} {\bibinfo {author} {\bibfnamefont {X.}~\bibnamefont
  {Zotos}}, \bibinfo {author} {\bibfnamefont {F.}~\bibnamefont {Naef}}, \ and\
  \bibinfo {author} {\bibfnamefont {P.}~\bibnamefont {Prelov\v{s}ek}},\ }\href
  {\doibase 10.1103/physrevb.55.11029} {\bibfield  {journal} {\bibinfo
  {journal} {Physical Review B}\ }\textbf {\bibinfo {volume} {55}},\ \bibinfo
  {pages} {11029} (\bibinfo {year} {1997})}\BibitemShut {NoStop}%
\bibitem [{\citenamefont {Ilievski}\ and\ \citenamefont {Prosen}(2012)}]{IP12}%
  \BibitemOpen
  \bibfield  {author} {\bibinfo {author} {\bibfnamefont {E.}~\bibnamefont
  {Ilievski}}\ and\ \bibinfo {author} {\bibfnamefont {T.}~\bibnamefont
  {Prosen}},\ }\href {\doibase 10.1007/s00220-012-1599-4} {\bibfield  {journal}
  {\bibinfo  {journal} {Communications in Mathematical Physics}\ }\textbf
  {\bibinfo {volume} {318}},\ \bibinfo {pages} {809} (\bibinfo {year}
  {2012})}\BibitemShut {NoStop}%
\bibitem [{\citenamefont {Yang}\ and\ \citenamefont {Yang}(1969)}]{YY69}%
  \BibitemOpen
  \bibfield  {author} {\bibinfo {author} {\bibfnamefont {C.-N.}\ \bibnamefont
  {Yang}}\ and\ \bibinfo {author} {\bibfnamefont {C.}~\bibnamefont {Yang}},\
  }\href {\doibase 10.1063/1.1664947} {\bibfield  {journal} {\bibinfo
  {journal} {Journal of Mathematical Physics}\ }\textbf {\bibinfo {volume}
  {10}},\ \bibinfo {pages} {1115} (\bibinfo {year} {1969})}\BibitemShut
  {NoStop}%
\bibitem [{\citenamefont {Takahashi}(1971)}]{Takahashi71}%
  \BibitemOpen
  \bibfield  {author} {\bibinfo {author} {\bibfnamefont {M.}~\bibnamefont
  {Takahashi}},\ }\href {\doibase 10.1143/ptp.46.401} {\bibfield  {journal}
  {\bibinfo  {journal} {Progress of Theoretical Physics}\ }\textbf {\bibinfo
  {volume} {46}},\ \bibinfo {pages} {401} (\bibinfo {year} {1971})}\BibitemShut
  {NoStop}%
\bibitem [{\citenamefont {Gaudin}(1971)}]{Gaudin71}%
  \BibitemOpen
  \bibfield  {author} {\bibinfo {author} {\bibfnamefont {M.}~\bibnamefont
  {Gaudin}},\ }\href {\doibase 10.1103/physrevlett.26.1301} {\bibfield
  {journal} {\bibinfo  {journal} {Physical Review Letters}\ }\textbf {\bibinfo
  {volume} {26}},\ \bibinfo {pages} {1301} (\bibinfo {year}
  {1971})}\BibitemShut {NoStop}%
\bibitem [{\citenamefont {Ilievski}\ \emph {et~al.}(2016)\citenamefont
  {Ilievski}, \citenamefont {Quinn}, \citenamefont {De~Nardis},\ and\
  \citenamefont {Brockmann}}]{StringCharge}%
  \BibitemOpen
  \bibfield  {author} {\bibinfo {author} {\bibfnamefont {E.}~\bibnamefont
  {Ilievski}}, \bibinfo {author} {\bibfnamefont {E.}~\bibnamefont {Quinn}},
  \bibinfo {author} {\bibfnamefont {J.}~\bibnamefont {De~Nardis}}, \ and\
  \bibinfo {author} {\bibfnamefont {M.}~\bibnamefont {Brockmann}},\ }\href
  {\doibase 10.1088/1742-5468/2016/06/063101} {\bibfield  {journal} {\bibinfo
  {journal} {Journal of Statistical Mechanics: Theory and Experiment}\ }\textbf
  {\bibinfo {volume} {2016}},\ \bibinfo {pages} {063101} (\bibinfo {year}
  {2016})}\BibitemShut {NoStop}%
\bibitem [{\citenamefont {Ilievski}\ \emph {et~al.}(2017)\citenamefont
  {Ilievski}, \citenamefont {Quinn},\ and\ \citenamefont {Caux}}]{IQC17}%
  \BibitemOpen
  \bibfield  {author} {\bibinfo {author} {\bibfnamefont {E.}~\bibnamefont
  {Ilievski}}, \bibinfo {author} {\bibfnamefont {E.}~\bibnamefont {Quinn}}, \
  and\ \bibinfo {author} {\bibfnamefont {J.-S.}\ \bibnamefont {Caux}},\ }\href
  {\doibase 10.1103/physrevb.95.115128} {\bibfield  {journal} {\bibinfo
  {journal} {Physical Review B}\ }\textbf {\bibinfo {volume} {95}} (\bibinfo
  {year} {2017}),\ 10.1103/physrevb.95.115128}\BibitemShut {NoStop}%
\bibitem [{\citenamefont {Spohn}\ and\ \citenamefont
  {Lebowitz}(1977)}]{Spohn77}%
  \BibitemOpen
  \bibfield  {author} {\bibinfo {author} {\bibfnamefont {H.}~\bibnamefont
  {Spohn}}\ and\ \bibinfo {author} {\bibfnamefont {J.~L.}\ \bibnamefont
  {Lebowitz}},\ }\href {\doibase 10.1007/bf01614132} {\bibfield  {journal}
  {\bibinfo  {journal} {Communications in Mathematical Physics}\ }\textbf
  {\bibinfo {volume} {54}},\ \bibinfo {pages} {97} (\bibinfo {year}
  {1977})}\BibitemShut {NoStop}%
\bibitem [{\citenamefont {Bernard}\ and\ \citenamefont {Doyon}(2014)}]{BD14}%
  \BibitemOpen
  \bibfield  {author} {\bibinfo {author} {\bibfnamefont {D.}~\bibnamefont
  {Bernard}}\ and\ \bibinfo {author} {\bibfnamefont {B.}~\bibnamefont
  {Doyon}},\ }\href {\doibase 10.1007/s00023-014-0314-8} {\bibfield  {journal}
  {\bibinfo  {journal} {Annales Henri Poincar{\'{e}}}\ }\textbf {\bibinfo
  {volume} {16}},\ \bibinfo {pages} {113} (\bibinfo {year} {2014})}\BibitemShut
  {NoStop}%
\bibitem [{\citenamefont {Bernard}\ and\ \citenamefont
  {Doyon}(2016)}]{BD16_review}%
  \BibitemOpen
  \bibfield  {author} {\bibinfo {author} {\bibfnamefont {D.}~\bibnamefont
  {Bernard}}\ and\ \bibinfo {author} {\bibfnamefont {B.}~\bibnamefont
  {Doyon}},\ }\href {\doibase 10.1088/1742-5468/2016/06/064005} {\bibfield
  {journal} {\bibinfo  {journal} {J. Stat. Mech.}\ }\textbf {\bibinfo {volume}
  {2016}},\ \bibinfo {pages} {064005} (\bibinfo {year} {2016})}\BibitemShut
  {NoStop}%
\bibitem [{\citenamefont {Karrasch}(2017{\natexlab{a}})}]{Karrasch17}%
  \BibitemOpen
  \bibfield  {author} {\bibinfo {author} {\bibfnamefont {C.}~\bibnamefont
  {Karrasch}},\ }\href {\doibase 10.1088/1367-2630/aa631a} {\bibfield
  {journal} {\bibinfo  {journal} {New Journal of Physics}\ }\textbf {\bibinfo
  {volume} {19}},\ \bibinfo {pages} {033027} (\bibinfo {year}
  {2017}{\natexlab{a}})}\BibitemShut {NoStop}%
\bibitem [{\citenamefont {Karrasch}\ \emph {et~al.}(2016)\citenamefont
  {Karrasch}, \citenamefont {Kennes},\ and\ \citenamefont
  {Heidrich-Meisner}}]{KKH16}%
  \BibitemOpen
  \bibfield  {author} {\bibinfo {author} {\bibfnamefont {C.}~\bibnamefont
  {Karrasch}}, \bibinfo {author} {\bibfnamefont {D.~M.}\ \bibnamefont
  {Kennes}}, \ and\ \bibinfo {author} {\bibfnamefont {F.}~\bibnamefont
  {Heidrich-Meisner}},\ }\href {\doibase 10.1103/physrevlett.117.116401}
  {\bibfield  {journal} {\bibinfo  {journal} {Physical Review Letters}\
  }\textbf {\bibinfo {volume} {117}} (\bibinfo {year} {2016}),\
  10.1103/physrevlett.117.116401}\BibitemShut {NoStop}%
\bibitem [{\citenamefont {Vasseur}\ \emph {et~al.}(2015)\citenamefont
  {Vasseur}, \citenamefont {Karrasch},\ and\ \citenamefont {Moore}}]{VKM16}%
  \BibitemOpen
  \bibfield  {author} {\bibinfo {author} {\bibfnamefont {R.}~\bibnamefont
  {Vasseur}}, \bibinfo {author} {\bibfnamefont {C.}~\bibnamefont {Karrasch}}, \
  and\ \bibinfo {author} {\bibfnamefont {J.~E.}\ \bibnamefont {Moore}},\ }\href
  {\doibase 10.1103/PhysRevLett.115.267201} {\bibfield  {journal} {\bibinfo
  {journal} {Physical Review Letters}\ }\textbf {\bibinfo {volume} {115}}
  (\bibinfo {year} {2015}),\ 10.1103/PhysRevLett.115.267201}\BibitemShut
  {NoStop}%
\bibitem [{\citenamefont {Bulchandani}\ \emph {et~al.}(2017)\citenamefont
  {Bulchandani}, \citenamefont {Vasseur}, \citenamefont {Karrasch},\ and\
  \citenamefont {Moore}}]{BVKM17}%
  \BibitemOpen
  \bibfield  {author} {\bibinfo {author} {\bibfnamefont {V.~B.}\ \bibnamefont
  {Bulchandani}}, \bibinfo {author} {\bibfnamefont {R.}~\bibnamefont
  {Vasseur}}, \bibinfo {author} {\bibfnamefont {C.}~\bibnamefont {Karrasch}}, \
  and\ \bibinfo {author} {\bibfnamefont {J.~E.}\ \bibnamefont {Moore}},\
  }\href@noop {} {\bibfield  {journal} {\bibinfo  {journal} {arXiv preprint
  arXiv:1702.06146}\ } (\bibinfo {year} {2017})}\BibitemShut {NoStop}%
\bibitem [{\citenamefont {Onsager}(1931)}]{Onsager}%
  \BibitemOpen
  \bibfield  {author} {\bibinfo {author} {\bibfnamefont {L.}~\bibnamefont
  {Onsager}},\ }\href {\doibase 10.1103/PhysRev.37.405} {\bibfield  {journal}
  {\bibinfo  {journal} {Phys. Rev.}\ }\textbf {\bibinfo {volume} {37}},\
  \bibinfo {pages} {405} (\bibinfo {year} {1931})}\BibitemShut {NoStop}%
\bibitem [{\citenamefont {Nardis}\ and\ \citenamefont
  {Panfil}(2016)}]{SciPostPhys.1.2.015}%
  \BibitemOpen
  \bibfield  {author} {\bibinfo {author} {\bibfnamefont {J.~D.}\ \bibnamefont
  {Nardis}}\ and\ \bibinfo {author} {\bibfnamefont {M.}~\bibnamefont
  {Panfil}},\ }\href {\doibase 10.21468/SciPostPhys.1.2.015} {\bibfield
  {journal} {\bibinfo  {journal} {SciPost Phys.}\ }\textbf {\bibinfo {volume}
  {1}},\ \bibinfo {pages} {015} (\bibinfo {year} {2016})}\BibitemShut {NoStop}%
\bibitem [{\citenamefont {De~Nardis}\ \emph {et~al.}(2017)\citenamefont
  {De~Nardis}, \citenamefont {Panfil}, \citenamefont {Gambassi}, \citenamefont
  {Cugliandolo}, \citenamefont {Konik},\ and\ \citenamefont
  {Foini}}]{FoiniDeNardis}%
  \BibitemOpen
  \bibfield  {author} {\bibinfo {author} {\bibfnamefont {J.}~\bibnamefont
  {De~Nardis}}, \bibinfo {author} {\bibfnamefont {M.}~\bibnamefont {Panfil}},
  \bibinfo {author} {\bibfnamefont {A.}~\bibnamefont {Gambassi}}, \bibinfo
  {author} {\bibfnamefont {L.~F.}\ \bibnamefont {Cugliandolo}}, \bibinfo
  {author} {\bibfnamefont {R.}~\bibnamefont {Konik}}, \ and\ \bibinfo {author}
  {\bibfnamefont {L.}~\bibnamefont {Foini}},\ }\href
  {https://arxiv.org/abs/1704.06649} {\bibfield  {journal} {\bibinfo  {journal}
  {arXiv preprint arXiv:1704.06649}\ } (\bibinfo {year} {2017})}\BibitemShut
  {NoStop}%
\bibitem [{\citenamefont {Gutzwiller}(1963)}]{Gutzwiller63}%
  \BibitemOpen
  \bibfield  {author} {\bibinfo {author} {\bibfnamefont {M.~C.}\ \bibnamefont
  {Gutzwiller}},\ }\href {\doibase 10.1103/physrevlett.10.159} {\bibfield
  {journal} {\bibinfo  {journal} {Physical Review Letters}\ }\textbf {\bibinfo
  {volume} {10}},\ \bibinfo {pages} {159} (\bibinfo {year} {1963})}\BibitemShut
  {NoStop}%
\bibitem [{\citenamefont {Hubbard}(1963)}]{Hubbard63}%
  \BibitemOpen
  \bibfield  {author} {\bibinfo {author} {\bibfnamefont {J.}~\bibnamefont
  {Hubbard}},\ }\href {\doibase 10.1098/rspa.1963.0204} {\bibfield  {journal}
  {\bibinfo  {journal} {Proceedings of the Royal Society A: Mathematical,
  Physical and Engineering Sciences}\ }\textbf {\bibinfo {volume} {276}},\
  \bibinfo {pages} {238} (\bibinfo {year} {1963})}\BibitemShut {NoStop}%
\bibitem [{\citenamefont {Shastry}(1988)}]{Shastry88}%
  \BibitemOpen
  \bibfield  {author} {\bibinfo {author} {\bibfnamefont {B.~S.}\ \bibnamefont
  {Shastry}},\ }\href {\doibase 10.1007/bf01022987} {\bibfield  {journal}
  {\bibinfo  {journal} {Journal of Statistical Physics}\ }\textbf {\bibinfo
  {volume} {50}},\ \bibinfo {pages} {57} (\bibinfo {year} {1988})}\BibitemShut
  {NoStop}%
\bibitem [{\citenamefont {Woynarovich}(1989)}]{Woynarovich89}%
  \BibitemOpen
  \bibfield  {author} {\bibinfo {author} {\bibfnamefont {F.}~\bibnamefont
  {Woynarovich}},\ }\href {\doibase 10.1088/0305-4470/22/19/017} {\bibfield
  {journal} {\bibinfo  {journal} {Journal of Physics A: Mathematical and
  General}\ }\textbf {\bibinfo {volume} {22}},\ \bibinfo {pages} {4243}
  (\bibinfo {year} {1989})}\BibitemShut {NoStop}%
\bibitem [{\citenamefont {Frahm}\ and\ \citenamefont {Korepin}(1990)}]{FK90}%
  \BibitemOpen
  \bibfield  {author} {\bibinfo {author} {\bibfnamefont {H.}~\bibnamefont
  {Frahm}}\ and\ \bibinfo {author} {\bibfnamefont {V.~E.}\ \bibnamefont
  {Korepin}},\ }\href {\doibase 10.1103/physrevb.42.10553} {\bibfield
  {journal} {\bibinfo  {journal} {Physical Review B}\ }\textbf {\bibinfo
  {volume} {42}},\ \bibinfo {pages} {10553} (\bibinfo {year}
  {1990})}\BibitemShut {NoStop}%
\bibitem [{\citenamefont {Ogata}\ and\ \citenamefont {Shiba}(1990)}]{OS90}%
  \BibitemOpen
  \bibfield  {author} {\bibinfo {author} {\bibfnamefont {M.}~\bibnamefont
  {Ogata}}\ and\ \bibinfo {author} {\bibfnamefont {H.}~\bibnamefont {Shiba}},\
  }\href {\doibase 10.1103/physrevb.41.2326} {\bibfield  {journal} {\bibinfo
  {journal} {Physical Review B}\ }\textbf {\bibinfo {volume} {41}},\ \bibinfo
  {pages} {2326} (\bibinfo {year} {1990})}\BibitemShut {NoStop}%
\bibitem [{\citenamefont {Prosen}\ and\ \citenamefont
  {{\v{Z}}nidari{\v{c}}}(2012)}]{PZ12}%
  \BibitemOpen
  \bibfield  {author} {\bibinfo {author} {\bibfnamefont {T.}~\bibnamefont
  {Prosen}}\ and\ \bibinfo {author} {\bibfnamefont {M.}~\bibnamefont
  {{\v{Z}}nidari{\v{c}}}},\ }\href {\doibase 10.1103/physrevb.86.125118}
  {\bibfield  {journal} {\bibinfo  {journal} {Physical Review B}\ }\textbf
  {\bibinfo {volume} {86}} (\bibinfo {year} {2012}),\
  10.1103/physrevb.86.125118}\BibitemShut {NoStop}%
\bibitem [{\citenamefont {Karrasch}\ \emph {et~al.}(2014)\citenamefont
  {Karrasch}, \citenamefont {Kennes},\ and\ \citenamefont {Moore}}]{KKM14}%
  \BibitemOpen
  \bibfield  {author} {\bibinfo {author} {\bibfnamefont {C.}~\bibnamefont
  {Karrasch}}, \bibinfo {author} {\bibfnamefont {D.~M.}\ \bibnamefont
  {Kennes}}, \ and\ \bibinfo {author} {\bibfnamefont {J.~E.}\ \bibnamefont
  {Moore}},\ }\href {\doibase 10.1103/physrevb.90.155104} {\bibfield  {journal}
  {\bibinfo  {journal} {Physical Review B}\ }\textbf {\bibinfo {volume} {90}}
  (\bibinfo {year} {2014}),\ 10.1103/physrevb.90.155104}\BibitemShut {NoStop}%
\bibitem [{\citenamefont {Prosen}(2014)}]{Prosen14}%
  \BibitemOpen
  \bibfield  {author} {\bibinfo {author} {\bibfnamefont {T.}~\bibnamefont
  {Prosen}},\ }\href {\doibase 10.1103/PhysRevLett.112.030603} {\bibfield
  {journal} {\bibinfo  {journal} {Phys. Rev. Lett.}\ }\textbf {\bibinfo
  {volume} {112}},\ \bibinfo {pages} {030603} (\bibinfo {year}
  {2014})}\BibitemShut {NoStop}%
\bibitem [{\citenamefont {Seabra}\ \emph {et~al.}(2014)\citenamefont {Seabra},
  \citenamefont {Essler}, \citenamefont {Pollmann}, \citenamefont {Schneider},\
  and\ \citenamefont {Veness}}]{Seabra14}%
  \BibitemOpen
  \bibfield  {author} {\bibinfo {author} {\bibfnamefont {L.}~\bibnamefont
  {Seabra}}, \bibinfo {author} {\bibfnamefont {F.~H.~L.}\ \bibnamefont
  {Essler}}, \bibinfo {author} {\bibfnamefont {F.}~\bibnamefont {Pollmann}},
  \bibinfo {author} {\bibfnamefont {I.}~\bibnamefont {Schneider}}, \ and\
  \bibinfo {author} {\bibfnamefont {T.}~\bibnamefont {Veness}},\ }\href
  {\doibase 10.1103/PhysRevB.90.245127} {\bibfield  {journal} {\bibinfo
  {journal} {Phys. Rev. B}\ }\textbf {\bibinfo {volume} {90}},\ \bibinfo
  {pages} {245127} (\bibinfo {year} {2014})}\BibitemShut {NoStop}%
\bibitem [{\citenamefont {Neumayer}\ \emph {et~al.}(2015)\citenamefont
  {Neumayer}, \citenamefont {Arrigoni}, \citenamefont {Aichhorn},\ and\
  \citenamefont {von~der Linden}}]{Neumayer15}%
  \BibitemOpen
  \bibfield  {author} {\bibinfo {author} {\bibfnamefont {J.}~\bibnamefont
  {Neumayer}}, \bibinfo {author} {\bibfnamefont {E.}~\bibnamefont {Arrigoni}},
  \bibinfo {author} {\bibfnamefont {M.}~\bibnamefont {Aichhorn}}, \ and\
  \bibinfo {author} {\bibfnamefont {W.}~\bibnamefont {von~der Linden}},\ }\href
  {\doibase 10.1103/PhysRevB.92.125149} {\bibfield  {journal} {\bibinfo
  {journal} {Phys. Rev. B}\ }\textbf {\bibinfo {volume} {92}},\ \bibinfo
  {pages} {125149} (\bibinfo {year} {2015})}\BibitemShut {NoStop}%
\bibitem [{\citenamefont {Reiner}\ \emph {et~al.}(2016)\citenamefont {Reiner},
  \citenamefont {Marthaler}, \citenamefont {Braum\"uller}, \citenamefont
  {Weides},\ and\ \citenamefont {Sch\"on}}]{Reiner16}%
  \BibitemOpen
  \bibfield  {author} {\bibinfo {author} {\bibfnamefont {J.-M.}\ \bibnamefont
  {Reiner}}, \bibinfo {author} {\bibfnamefont {M.}~\bibnamefont {Marthaler}},
  \bibinfo {author} {\bibfnamefont {J.}~\bibnamefont {Braum\"uller}}, \bibinfo
  {author} {\bibfnamefont {M.}~\bibnamefont {Weides}}, \ and\ \bibinfo {author}
  {\bibfnamefont {G.}~\bibnamefont {Sch\"on}},\ }\href {\doibase
  10.1103/PhysRevA.94.032338} {\bibfield  {journal} {\bibinfo  {journal} {Phys.
  Rev. A}\ }\textbf {\bibinfo {volume} {94}},\ \bibinfo {pages} {032338}
  (\bibinfo {year} {2016})}\BibitemShut {NoStop}%
\bibitem [{\citenamefont {Veness}\ and\ \citenamefont {Essler}(2016)}]{VE16}%
  \BibitemOpen
  \bibfield  {author} {\bibinfo {author} {\bibfnamefont {T.}~\bibnamefont
  {Veness}}\ and\ \bibinfo {author} {\bibfnamefont {F.~H.~L.}\ \bibnamefont
  {Essler}},\ }\href {\doibase 10.1103/PhysRevB.93.205101} {\bibfield
  {journal} {\bibinfo  {journal} {Phys. Rev. B}\ }\textbf {\bibinfo {volume}
  {93}},\ \bibinfo {pages} {205101} (\bibinfo {year} {2016})}\BibitemShut
  {NoStop}%
\bibitem [{\citenamefont {Veness}\ \emph {et~al.}(2016)\citenamefont {Veness},
  \citenamefont {Essler},\ and\ \citenamefont {Fisher}}]{Vaness16}%
  \BibitemOpen
  \bibfield  {author} {\bibinfo {author} {\bibfnamefont {T.}~\bibnamefont
  {Veness}}, \bibinfo {author} {\bibfnamefont {F.~H.~L.}\ \bibnamefont
  {Essler}}, \ and\ \bibinfo {author} {\bibfnamefont {M.~P.~A.}\ \bibnamefont
  {Fisher}},\ }\href {{https://arxiv.org/abs/1611.02075}} {\bibfield  {journal}
  {\bibinfo  {journal} {arXiv preprint arXiv:1611.02075}\ } (\bibinfo {year}
  {2016})}\BibitemShut {NoStop}%
\bibitem [{\citenamefont {Tiegel}\ \emph {et~al.}(2016)\citenamefont {Tiegel},
  \citenamefont {Veness}, \citenamefont {Dargel}, \citenamefont {Honecker},
  \citenamefont {Pruschke}, \citenamefont {McCulloch},\ and\ \citenamefont
  {Essler}}]{Tiegel16}%
  \BibitemOpen
  \bibfield  {author} {\bibinfo {author} {\bibfnamefont {A.~C.}\ \bibnamefont
  {Tiegel}}, \bibinfo {author} {\bibfnamefont {T.}~\bibnamefont {Veness}},
  \bibinfo {author} {\bibfnamefont {P.~E.}\ \bibnamefont {Dargel}}, \bibinfo
  {author} {\bibfnamefont {A.}~\bibnamefont {Honecker}}, \bibinfo {author}
  {\bibfnamefont {T.}~\bibnamefont {Pruschke}}, \bibinfo {author}
  {\bibfnamefont {I.~P.}\ \bibnamefont {McCulloch}}, \ and\ \bibinfo {author}
  {\bibfnamefont {F.~H.~L.}\ \bibnamefont {Essler}},\ }\href {\doibase
  10.1103/PhysRevB.93.125108} {\bibfield  {journal} {\bibinfo  {journal} {Phys.
  Rev. B}\ }\textbf {\bibinfo {volume} {93}},\ \bibinfo {pages} {125108}
  (\bibinfo {year} {2016})}\BibitemShut {NoStop}%
\bibitem [{\citenamefont {Karrasch}\ \emph {et~al.}(2017)\citenamefont
  {Karrasch}, \citenamefont {Prosen},\ and\ \citenamefont
  {Heidrich-Meisner}}]{KPH17}%
  \BibitemOpen
  \bibfield  {author} {\bibinfo {author} {\bibfnamefont {C.}~\bibnamefont
  {Karrasch}}, \bibinfo {author} {\bibfnamefont {T.}~\bibnamefont {Prosen}}, \
  and\ \bibinfo {author} {\bibfnamefont {F.}~\bibnamefont {Heidrich-Meisner}},\
  }\href {\doibase 10.1103/physrevb.95.060406} {\bibfield  {journal} {\bibinfo
  {journal} {Physical Review B}\ }\textbf {\bibinfo {volume} {95}} (\bibinfo
  {year} {2017}),\ 10.1103/physrevb.95.060406}\BibitemShut {NoStop}%
\bibitem [{\citenamefont {Lin}\ \emph {et~al.}(2017)\citenamefont {Lin},
  \citenamefont {Huang},\ and\ \citenamefont {Hin}}]{Lin17}%
  \BibitemOpen
  \bibfield  {author} {\bibinfo {author} {\bibfnamefont {F.}~\bibnamefont
  {Lin}}, \bibinfo {author} {\bibfnamefont {J.}~\bibnamefont {Huang}}, \ and\
  \bibinfo {author} {\bibfnamefont {C.}~\bibnamefont {Hin}},\ }\href
  {{https://arxiv.org/abs/1704.07545}} {\bibfield  {journal} {\bibinfo
  {journal} {arXiv preprint arXiv:1704.07545}\ } (\bibinfo {year}
  {2017})}\BibitemShut {NoStop}%
\bibitem [{\citenamefont {Karrasch}(2017{\natexlab{b}})}]{Karrasch17HUB}%
  \BibitemOpen
  \bibfield  {author} {\bibinfo {author} {\bibfnamefont {C.}~\bibnamefont
  {Karrasch}},\ }\href {\doibase 10.1103/PhysRevB.95.115148} {\bibfield
  {journal} {\bibinfo  {journal} {Phys. Rev. B}\ }\textbf {\bibinfo {volume}
  {95}},\ \bibinfo {pages} {115148} (\bibinfo {year}
  {2017}{\natexlab{b}})}\BibitemShut {NoStop}%
\bibitem [{\citenamefont {Bolens}\ \emph {et~al.}(2017)\citenamefont {Bolens},
  \citenamefont {Katsura}, \citenamefont {Ogata},\ and\ \citenamefont
  {Miyashita}}]{Bolens17}%
  \BibitemOpen
  \bibfield  {author} {\bibinfo {author} {\bibfnamefont {A.}~\bibnamefont
  {Bolens}}, \bibinfo {author} {\bibfnamefont {H.}~\bibnamefont {Katsura}},
  \bibinfo {author} {\bibfnamefont {M.}~\bibnamefont {Ogata}}, \ and\ \bibinfo
  {author} {\bibfnamefont {S.}~\bibnamefont {Miyashita}},\ }\href {\doibase
  10.1103/PhysRevB.95.235115} {\bibfield  {journal} {\bibinfo  {journal} {Phys.
  Rev. B}\ }\textbf {\bibinfo {volume} {95}},\ \bibinfo {pages} {235115}
  (\bibinfo {year} {2017})}\BibitemShut {NoStop}%
\bibitem [{\citenamefont {Lieb}\ and\ \citenamefont {Wu}(1968)}]{LW68}%
  \BibitemOpen
  \bibfield  {author} {\bibinfo {author} {\bibfnamefont {E.~H.}\ \bibnamefont
  {Lieb}}\ and\ \bibinfo {author} {\bibfnamefont {F.~Y.}\ \bibnamefont {Wu}},\
  }\href {\doibase 10.1103/physrevlett.21.192.2} {\bibfield  {journal}
  {\bibinfo  {journal} {Physical Review Letters}\ }\textbf {\bibinfo {volume}
  {21}},\ \bibinfo {pages} {192} (\bibinfo {year} {1968})}\BibitemShut
  {NoStop}%
\bibitem [{\citenamefont {Takahashi}(1972)}]{Takahashi_Hubbard}%
  \BibitemOpen
  \bibfield  {author} {\bibinfo {author} {\bibfnamefont {M.}~\bibnamefont
  {Takahashi}},\ }\href {\doibase 10.1143/ptp.47.69} {\bibfield  {journal}
  {\bibinfo  {journal} {Progress of Theoretical Physics}\ }\textbf {\bibinfo
  {volume} {47}},\ \bibinfo {pages} {69} (\bibinfo {year} {1972})}\BibitemShut
  {NoStop}%
\bibitem [{\citenamefont {Prosen}(2011)}]{Prosen11}%
  \BibitemOpen
  \bibfield  {author} {\bibinfo {author} {\bibfnamefont {T.}~\bibnamefont
  {Prosen}},\ }\href {\doibase 10.1103/physrevlett.106.217206} {\bibfield
  {journal} {\bibinfo  {journal} {Physical Review Letters}\ }\textbf {\bibinfo
  {volume} {106}} (\bibinfo {year} {2011}),\
  10.1103/physrevlett.106.217206}\BibitemShut {NoStop}%
\bibitem [{\citenamefont {Prosen}\ and\ \citenamefont {Ilievski}(2013)}]{PI13}%
  \BibitemOpen
  \bibfield  {author} {\bibinfo {author} {\bibfnamefont {T.}~\bibnamefont
  {Prosen}}\ and\ \bibinfo {author} {\bibfnamefont {E.}~\bibnamefont
  {Ilievski}},\ }\href {\doibase 10.1103/physrevlett.111.057203} {\bibfield
  {journal} {\bibinfo  {journal} {Physical Review Letters}\ }\textbf {\bibinfo
  {volume} {111}} (\bibinfo {year} {2013}),\
  10.1103/physrevlett.111.057203}\BibitemShut {NoStop}%
\bibitem [{\citenamefont {Shastry}\ and\ \citenamefont
  {Sutherland}(1990)}]{SS90}%
  \BibitemOpen
  \bibfield  {author} {\bibinfo {author} {\bibfnamefont {B.~S.}\ \bibnamefont
  {Shastry}}\ and\ \bibinfo {author} {\bibfnamefont {B.}~\bibnamefont
  {Sutherland}},\ }\href {\doibase 10.1103/physrevlett.65.243} {\bibfield
  {journal} {\bibinfo  {journal} {Physical Review Letters}\ }\textbf {\bibinfo
  {volume} {65}},\ \bibinfo {pages} {243} (\bibinfo {year} {1990})}\BibitemShut
  {NoStop}%
\bibitem [{\citenamefont {Fujimoto}\ and\ \citenamefont
  {Kawakami}(1998)}]{FK98}%
  \BibitemOpen
  \bibfield  {author} {\bibinfo {author} {\bibfnamefont {S.}~\bibnamefont
  {Fujimoto}}\ and\ \bibinfo {author} {\bibfnamefont {N.}~\bibnamefont
  {Kawakami}},\ }\href {\doibase 10.1088/0305-4470/31/2/008} {\bibfield
  {journal} {\bibinfo  {journal} {{Journal of Physics A: Mathematical and
  General}}\ }\textbf {\bibinfo {volume} {31}},\ \bibinfo {pages} {465}
  (\bibinfo {year} {1998})}\BibitemShut {NoStop}%
\bibitem [{\citenamefont {Zotos}(1999)}]{Zotos99}%
  \BibitemOpen
  \bibfield  {author} {\bibinfo {author} {\bibfnamefont {X.}~\bibnamefont
  {Zotos}},\ }\href {\doibase 10.1103/physrevlett.82.1764} {\bibfield
  {journal} {\bibinfo  {journal} {Physical Review Letters}\ }\textbf {\bibinfo
  {volume} {82}},\ \bibinfo {pages} {1764} (\bibinfo {year}
  {1999})}\BibitemShut {NoStop}%
\bibitem [{\citenamefont {Benz}\ \emph {et~al.}(2005)\citenamefont {Benz},
  \citenamefont {Fukui}, \citenamefont {Kl\"{u}mper},\ and\ \citenamefont
  {Scheeren}}]{BFKS05}%
  \BibitemOpen
  \bibfield  {author} {\bibinfo {author} {\bibfnamefont {J.}~\bibnamefont
  {Benz}}, \bibinfo {author} {\bibfnamefont {T.}~\bibnamefont {Fukui}},
  \bibinfo {author} {\bibfnamefont {A.}~\bibnamefont {Kl\"{u}mper}}, \ and\
  \bibinfo {author} {\bibfnamefont {C.}~\bibnamefont {Scheeren}},\ }\href
  {\doibase 10.1143/jpsjs.74s.181} {\bibfield  {journal} {\bibinfo  {journal}
  {Journal of the Physical Society of Japan}\ }\textbf {\bibinfo {volume}
  {74}},\ \bibinfo {pages} {181} (\bibinfo {year} {2005})}\BibitemShut
  {NoStop}%
\bibitem [{\citenamefont {Kohn}(1964)}]{Kohn64}%
  \BibitemOpen
  \bibfield  {author} {\bibinfo {author} {\bibfnamefont {W.}~\bibnamefont
  {Kohn}},\ }\href {\doibase 10.1103/physrev.133.a171} {\bibfield  {journal}
  {\bibinfo  {journal} {Physical Review}\ }\textbf {\bibinfo {volume} {133}},\
  \bibinfo {pages} {A171} (\bibinfo {year} {1964})}\BibitemShut {NoStop}%
\bibitem [{\citenamefont {Narozhny}\ \emph {et~al.}(1998)\citenamefont
  {Narozhny}, \citenamefont {Millis},\ and\ \citenamefont
  {Andrei}}]{Narozhny98}%
  \BibitemOpen
  \bibfield  {author} {\bibinfo {author} {\bibfnamefont {B.~N.}\ \bibnamefont
  {Narozhny}}, \bibinfo {author} {\bibfnamefont {A.~J.}\ \bibnamefont
  {Millis}}, \ and\ \bibinfo {author} {\bibfnamefont {N.}~\bibnamefont
  {Andrei}},\ }\href {\doibase 10.1103/physrevb.58.r2921} {\bibfield  {journal}
  {\bibinfo  {journal} {Physical Review B}\ }\textbf {\bibinfo {volume} {58}},\
  \bibinfo {pages} {R2921} (\bibinfo {year} {1998})}\BibitemShut {NoStop}%
\bibitem [{\citenamefont {Peres}\ \emph {et~al.}(1999)\citenamefont {Peres},
  \citenamefont {Sacramento}, \citenamefont {Campbell},\ and\ \citenamefont
  {Carmelo}}]{Peres99}%
  \BibitemOpen
  \bibfield  {author} {\bibinfo {author} {\bibfnamefont {N.~M.~R.}\
  \bibnamefont {Peres}}, \bibinfo {author} {\bibfnamefont {P.~D.}\ \bibnamefont
  {Sacramento}}, \bibinfo {author} {\bibfnamefont {D.~K.}\ \bibnamefont
  {Campbell}}, \ and\ \bibinfo {author} {\bibfnamefont {J.~M.~P.}\ \bibnamefont
  {Carmelo}},\ }\href {\doibase 10.1103/physrevb.59.7382} {\bibfield  {journal}
  {\bibinfo  {journal} {Physical Review B}\ }\textbf {\bibinfo {volume} {59}},\
  \bibinfo {pages} {7382} (\bibinfo {year} {1999})}\BibitemShut {NoStop}%
\bibitem [{\citenamefont {Alvarez}\ and\ \citenamefont {Gros}(2002)}]{AG02}%
  \BibitemOpen
  \bibfield  {author} {\bibinfo {author} {\bibfnamefont {J.~V.}\ \bibnamefont
  {Alvarez}}\ and\ \bibinfo {author} {\bibfnamefont {C.}~\bibnamefont {Gros}},\
  }\href {\doibase 10.1103/physrevlett.88.077203} {\bibfield  {journal}
  {\bibinfo  {journal} {Physical Review Letters}\ }\textbf {\bibinfo {volume}
  {88}} (\bibinfo {year} {2002}),\ 10.1103/physrevlett.88.077203}\BibitemShut
  {NoStop}%
\bibitem [{\citenamefont {Heidrich-Meisner}\ \emph {et~al.}(2003)\citenamefont
  {Heidrich-Meisner}, \citenamefont {Honecker}, \citenamefont {Cabra},\ and\
  \citenamefont {Brenig}}]{HM03}%
  \BibitemOpen
  \bibfield  {author} {\bibinfo {author} {\bibfnamefont {F.}~\bibnamefont
  {Heidrich-Meisner}}, \bibinfo {author} {\bibfnamefont {A.}~\bibnamefont
  {Honecker}}, \bibinfo {author} {\bibfnamefont {D.~C.}\ \bibnamefont {Cabra}},
  \ and\ \bibinfo {author} {\bibfnamefont {W.}~\bibnamefont {Brenig}},\ }\href
  {\doibase 10.1103/physrevb.68.189901} {\bibfield  {journal} {\bibinfo
  {journal} {Physical Review B}\ }\textbf {\bibinfo {volume} {68}} (\bibinfo
  {year} {2003}),\ 10.1103/physrevb.68.189901}\BibitemShut {NoStop}%
\bibitem [{\citenamefont {Fujimoto}\ and\ \citenamefont
  {Kawakami}(2003)}]{FK03}%
  \BibitemOpen
  \bibfield  {author} {\bibinfo {author} {\bibfnamefont {S.}~\bibnamefont
  {Fujimoto}}\ and\ \bibinfo {author} {\bibfnamefont {N.}~\bibnamefont
  {Kawakami}},\ }\href {\doibase 10.1103/physrevlett.90.197202} {\bibfield
  {journal} {\bibinfo  {journal} {Physical Review Letters}\ }\textbf {\bibinfo
  {volume} {90}} (\bibinfo {year} {2003}),\
  10.1103/physrevlett.90.197202}\BibitemShut {NoStop}%
\bibitem [{\citenamefont {Herbrych}\ \emph {et~al.}(2011)\citenamefont
  {Herbrych}, \citenamefont {Prelov{\v{s}}ek},\ and\ \citenamefont
  {Zotos}}]{Herbrych11}%
  \BibitemOpen
  \bibfield  {author} {\bibinfo {author} {\bibfnamefont {J.}~\bibnamefont
  {Herbrych}}, \bibinfo {author} {\bibfnamefont {P.}~\bibnamefont
  {Prelov{\v{s}}ek}}, \ and\ \bibinfo {author} {\bibfnamefont {X.}~\bibnamefont
  {Zotos}},\ }\href {\doibase 10.1103/physrevb.84.155125} {\bibfield  {journal}
  {\bibinfo  {journal} {Physical Review B}\ }\textbf {\bibinfo {volume} {84}}
  (\bibinfo {year} {2011}),\ 10.1103/physrevb.84.155125}\BibitemShut {NoStop}%
\bibitem [{\citenamefont {Sirker}\ \emph {et~al.}(2011)\citenamefont {Sirker},
  \citenamefont {Pereira},\ and\ \citenamefont {Affleck}}]{SPA11}%
  \BibitemOpen
  \bibfield  {author} {\bibinfo {author} {\bibfnamefont {J.}~\bibnamefont
  {Sirker}}, \bibinfo {author} {\bibfnamefont {R.~G.}\ \bibnamefont {Pereira}},
  \ and\ \bibinfo {author} {\bibfnamefont {I.}~\bibnamefont {Affleck}},\ }\href
  {\doibase 10.1103/physrevb.83.035115} {\bibfield  {journal} {\bibinfo
  {journal} {Physical Review B}\ }\textbf {\bibinfo {volume} {83}} (\bibinfo
  {year} {2011}),\ 10.1103/physrevb.83.035115}\BibitemShut {NoStop}%
\bibitem [{\citenamefont {{\v{Z}}nidari{\v{c}}}(2011)}]{Znidaric11}%
  \BibitemOpen
  \bibfield  {author} {\bibinfo {author} {\bibfnamefont {M.}~\bibnamefont
  {{\v{Z}}nidari{\v{c}}}},\ }\href {\doibase 10.1103/physrevlett.106.220601}
  {\bibfield  {journal} {\bibinfo  {journal} {Physical Review Letters}\
  }\textbf {\bibinfo {volume} {106}} (\bibinfo {year} {2011}),\
  10.1103/physrevlett.106.220601}\BibitemShut {NoStop}%
\bibitem [{\citenamefont {Gromov}\ and\ \citenamefont
  {Kazakov}(2011)}]{Gromov11}%
  \BibitemOpen
  \bibfield  {author} {\bibinfo {author} {\bibfnamefont {N.}~\bibnamefont
  {Gromov}}\ and\ \bibinfo {author} {\bibfnamefont {V.}~\bibnamefont
  {Kazakov}},\ }\href {\doibase 10.1007/s11005-011-0513-x} {\bibfield
  {journal} {\bibinfo  {journal} {Letters in Mathematical Physics}\ }\textbf
  {\bibinfo {volume} {99}},\ \bibinfo {pages} {321} (\bibinfo {year}
  {2011})}\BibitemShut {NoStop}%
\bibitem [{\citenamefont {Karrasch}\ \emph {et~al.}(2012)\citenamefont
  {Karrasch}, \citenamefont {Bardarson},\ and\ \citenamefont {Moore}}]{KBM12}%
  \BibitemOpen
  \bibfield  {author} {\bibinfo {author} {\bibfnamefont {C.}~\bibnamefont
  {Karrasch}}, \bibinfo {author} {\bibfnamefont {J.~H.}\ \bibnamefont
  {Bardarson}}, \ and\ \bibinfo {author} {\bibfnamefont {J.~E.}\ \bibnamefont
  {Moore}},\ }\href {\doibase 10.1103/physrevlett.108.227206} {\bibfield
  {journal} {\bibinfo  {journal} {Physical Review Letters}\ }\textbf {\bibinfo
  {volume} {108}} (\bibinfo {year} {2012}),\
  10.1103/physrevlett.108.227206}\BibitemShut {NoStop}%
\bibitem [{\citenamefont {Cavagli{\`{a}}}\ \emph {et~al.}(2015)\citenamefont
  {Cavagli{\`{a}}}, \citenamefont {Cornagliotto}, \citenamefont {Mattelliano},\
  and\ \citenamefont {Tateo}}]{Cavaglia15}%
  \BibitemOpen
  \bibfield  {author} {\bibinfo {author} {\bibfnamefont {A.}~\bibnamefont
  {Cavagli{\`{a}}}}, \bibinfo {author} {\bibfnamefont {M.}~\bibnamefont
  {Cornagliotto}}, \bibinfo {author} {\bibfnamefont {M.}~\bibnamefont
  {Mattelliano}}, \ and\ \bibinfo {author} {\bibfnamefont {R.}~\bibnamefont
  {Tateo}},\ }\href {\doibase 10.1007/jhep06(2015)015} {\bibfield  {journal}
  {\bibinfo  {journal} {Journal of High Energy Physics}\ }\textbf {\bibinfo
  {volume} {2015}} (\bibinfo {year} {2015}),\
  10.1007/jhep06(2015)015}\BibitemShut {NoStop}%
\bibitem [{\citenamefont {De~Luca}\ \emph {et~al.}(2016)\citenamefont
  {De~Luca}, \citenamefont {Collura},\ and\ \citenamefont
  {De~Nardis}}]{DeLuca16}%
  \BibitemOpen
  \bibfield  {author} {\bibinfo {author} {\bibfnamefont {A.}~\bibnamefont
  {De~Luca}}, \bibinfo {author} {\bibfnamefont {M.}~\bibnamefont {Collura}}, \
  and\ \bibinfo {author} {\bibfnamefont {J.}~\bibnamefont {De~Nardis}},\
  }\href@noop {} {\bibfield  {journal} {\bibinfo  {journal} {arXiv preprint
  arXiv:1612.07265}\ } (\bibinfo {year} {2016})}\BibitemShut {NoStop}%
\bibitem [{\citenamefont {Foini}\ \emph {et~al.}(2017)\citenamefont {Foini},
  \citenamefont {Gambassi}, \citenamefont {Konik},\ and\ \citenamefont
  {Cugliandolo}}]{PhysRevE.95.052116}%
  \BibitemOpen
  \bibfield  {author} {\bibinfo {author} {\bibfnamefont {L.}~\bibnamefont
  {Foini}}, \bibinfo {author} {\bibfnamefont {A.}~\bibnamefont {Gambassi}},
  \bibinfo {author} {\bibfnamefont {R.}~\bibnamefont {Konik}}, \ and\ \bibinfo
  {author} {\bibfnamefont {L.~F.}\ \bibnamefont {Cugliandolo}},\ }\href
  {\doibase 10.1103/PhysRevE.95.052116} {\bibfield  {journal} {\bibinfo
  {journal} {Phys. Rev. E}\ }\textbf {\bibinfo {volume} {95}},\ \bibinfo
  {pages} {052116} (\bibinfo {year} {2017})}\BibitemShut {NoStop}%
\bibitem [{\citenamefont {Medenjak}\ \emph
  {et~al.}(2017{\natexlab{a}})\citenamefont {Medenjak}, \citenamefont
  {Karrasch},\ and\ \citenamefont {Prosen}}]{MKP17}%
  \BibitemOpen
  \bibfield  {author} {\bibinfo {author} {\bibfnamefont {M.}~\bibnamefont
  {Medenjak}}, \bibinfo {author} {\bibfnamefont {C.}~\bibnamefont {Karrasch}},
  \ and\ \bibinfo {author} {\bibfnamefont {T.}~\bibnamefont {Prosen}},\
  }\href@noop {} {\bibfield  {journal} {\bibinfo  {journal} {arXiv preprint
  arXiv:1702.04677}\ } (\bibinfo {year} {2017}{\natexlab{a}})}\BibitemShut
  {NoStop}%
\bibitem [{\citenamefont {Medenjak}\ \emph
  {et~al.}(2017{\natexlab{b}})\citenamefont {Medenjak}, \citenamefont
  {Klobas},\ and\ \citenamefont {Prosen}}]{Medenjak17}%
  \BibitemOpen
  \bibfield  {author} {\bibinfo {author} {\bibfnamefont {M.}~\bibnamefont
  {Medenjak}}, \bibinfo {author} {\bibfnamefont {K.}~\bibnamefont {Klobas}}, \
  and\ \bibinfo {author} {\bibfnamefont {T.}~\bibnamefont {Prosen}},\
  }\href@noop {} {\bibfield  {journal} {\bibinfo  {journal} {arXiv preprint
  arXiv:1705.04636}\ } (\bibinfo {year} {2017}{\natexlab{b}})}\BibitemShut
  {NoStop}%
\end{thebibliography}%



\pagebreak

\appendix
\onecolumngrid

\clearpage

\begin{center}
\textbf{{\Large Supplemental Material}}
\end{center}
\begin{center}
\textbf{{\large Ballistic transport in the one-dimensional Hubbard model:\\ the hydrodynamic approach}}
\end{center}

In this Supplementary Material we collect the most important technical results, present the detailed derivations and
provide additional numerical results. The structure is as follows:

\begin{itemize}

\item Appendix \ref{app:Hubbard_BA} covers the technical background of the nested Bethe Ansatz technique for solving the
one-dimensional Hubbard model. We follow closely the presentation of \cite{FQ12,QF13} which employs rapidity parametrization.
A quasi-local formulation of TBA equations and the dressing transformation presented here appear to be new.

\item In Appendix \ref{app:covariances} we give a short derivation for the full generalized charge-charge and
charge-current covariance matrices. This extends recent results of \cite{DS17} to integrable quantum models solvable
by nested Bethe Ansatz.

\item In Appendix \ref{app:linearized} we linearize the nonequilibrium hydrodynamic equations around a reference local
equilibrium state, and obtain analytic closed-form expression of the (generalized) Drude weights.
An analogous result for the Lieb--Liniger gas already appears in \cite{DS17}.

\item In Appendix \ref{app:detailed} we present a derivation of the generalized detailed balance condition for
an integrable model with multiple particle species by repeating the steps of the recent study \cite{SciPostPhys.1.2.015}
on the Lieb--Liniger model.

\item In Appendix \ref{app:Kohn} we briefly revisit the exceptional case of spin Drude weight in the anisotropic
Heisenberg spin-$1/2$ chain. We explain how to unify the three apriori different definitions for computing the spin Drude weigh
employ in the previous literature.

\item In Appendix \ref{app:hydro_Hubbard} we present a general solution to the hydrodynamic equations for Hubbard model
for the evolution from a bipartite initial equilibrium state. As an example, we compute the energy
density and energy current quasi-stationary profiles and compare them with the results of tDMRG simulation.

\end{itemize}

\section{Thermodynamic Bethe Ansatz for Hubbard model}
\label{app:Hubbard_BA}

The Hamiltonian of 1D Hubbard model is of the form
\begin{equation}
\hat{H}_{0} = -\sum_{x=1}^{L}\sum_{\sigma=\ua,\da}\Big(\hat{c}^{\dagger}_{x,\sigma}\hat{c}_{x+1,\sigma} +
\hat{c}^{\dagger}_{x+1,\sigma}\hat{c}_{x,\sigma}\Big)
+4\uu \sum_{x=1}^{L}(\hat{n}_{x,\ua}-\tfrac{1}{2})(\hat{n}_{x,\da}-\tfrac{1}{2}),
\label{eqn:H0}
\end{equation}
The model possesses two globally conserved charges associated with $U(1)$ symmetries, the total electron
charge $\hat{N}$ and the total spin $\hat{S}^{z}$,
\begin{equation}
\hat{N} = \sum_{x=1}^{L}\Big(\hat{c}^{\dagger}_{x,\ua}\hat{c}_{x,\ua} + \hat{c}^{\dagger}_{x,\da}\hat{c}_{x,\da}\Big),\qquad
\hat{S}^{z} = \frac{1}{2}\sum_{x=1}^{L}\Big(\hat{c}^{\dagger}_{x,\ua}\hat{c}_{x,\ua} - \hat{c}^{\dagger}_{x,\da}\hat{c}_{x,\da}\Big).
\end{equation}
which are sometimes included in the definition Hamiltonian, $\hat{H} = \hat{H}_{0} + \mu_{\rm c}\,\hat{N} + \mu_{\rm s}\,\hat{S}^{z}$.

Bethe equations for a finite system of length $L$ with periodic boundary conditions have been derived
by Lieb and Wu~\cite{LW68} and take the \emph{nested form},
\begin{align}
e^{\ii \p(u_{k})L}\prod_{j=1}^{M}\frac{u_{k}-w_{j}-\ii \uu}{u_{k}-w_{j}+\ii \uu} &= 1,\\
\prod_{j=1}^{N}\frac{w_{k}-v_{j}-\ii \uu}{w_{k}-u_{j}+\ii \uu}\prod_{m=1}^{M}\frac{w_{l}-w_{m}+2\ii \uu}{w_{l}-w_{m}-2\ii \uu} &= -1,
\label{eqn:LiebWu}
\end{align}
with $2M\leq N \leq L$. Bethe roots (rapidities) $u_{k}$ are related to electron (quasi)momenta, while $w_{k}$ are 
associated with their spin. The number of Bethe roots in Eqs.~\eqref{eqn:LiebWu} in terms of the total charge and spin is $N$ and 
$(N-2M)/2$, respectively. Bethe roots are associated the bare charge $n_{u}=1$, $n_{w}=0$, and the bare spin $m_{u}=\tfrac{1}{2}$, $m_{w}=-1$. We note that parametrization of Lieb--Wu equations \eqref{eqn:LiebWu} in terms of $u$-roots is different from the
conventional one given in terms of electron (quasi)momenta $p_{j}$ as in \cite{Takahashi_Hubbard}.
While the two are simply related by $u_{j}=\sin{(\p_{j})}$, rapidity parametrization proves more convenient since it renders
all scattering amplitudes manifestly rational functions depending only on the difference of particles' rapidities.
A downside is that the momentum-dependent phase $e^{\ii \p(u)}$ as a function of momentum-carrying roots $u_{i}$ then
becomes a double-valued function, meaning that each root $u_{i}$ gives two distinct values of momenta.
It is thus convenient to introduce a new type of roots, referred to as the $y$-roots, by virtue of Zhukovsky transform
\begin{equation}
u_{i} = \frac{1}{2}\left(y_{i}+\frac{1}{y_{i}}\right).
\label{eqn:Zhukovsky}
\end{equation}
The corresponding functional equation $\frac{1}{2}\left(x(u)+1/x(u)\right) = u$ has two solutions (branches),
and presently we adopt
\begin{equation}
x(u) = u + u\sqrt{1-\frac{1}{u^{2}}},
\label{eqn:function_x}
\end{equation}
with a square-root branch cut on the interval $\mathcal{I}=(-1,1)$.
For any $u\in \mathbb{C}$ the two branches correspond to the values $y_{\pm}(u)$ given by
\begin{equation}
y_{+}(u) = x(u),\qquad y_{-}(u) = \frac{1}{x(u)}.
\end{equation}
When rapidity $u$ is taken from the branch cut, $u\in \mathcal{I}$, we adopt the following prescription
\begin{equation}
y_{\pm}(u) = x(u\pm \ii 0),
\end{equation}
i.e. we take the two values just above and just below the cut $\mathcal{I}$. Since we have
\begin{equation}
e^{\ii \p(u)} = \ii y,
\end{equation}
the two branches of momenta are given by
\begin{equation}
\p_{\pm}(u) = -\ii \log(\ii\,x(u\pm \ii 0)).
\end{equation}

Thermodynamic solutions to Eqs.~\eqref{eqn:LiebWu} -- taking the limit $L\to \infty$ while keeping ratios 
$N/L$ and $M/L$ finite -- can be inferred from the stability condition of the asymptotic solutions, and comprise of
self-conjugate string-like patterns of regularly displaced complex-valued rapidities with equal real parts centred on the real axis.
These are identified with the thermodynamic particle content of the model which in the Hubbard model and comprise of:
\begin{itemize}
\item \underline{$y$-particles}, which are spin-up $m_{y}=\tfrac{1}{2}$ momentum-carrying electronic excitations which carry unit 
electron charge $n_{y}=1$. The $y$-particle excitations are split into two branches denoted by $y_{\pm}$ with the
corresponding rapidities $u_{\pm j}$. The $y$-particles do not form bound states on their own.
Their bare momenta are denoted by $\p_{\pm}(u)\in \mathbb{R}$, and satisfy $\p^{\prime}_{-}>0$ and $\p^{\prime}_{+}<0$,
with $\p^{\prime}_{+} + \p^{\prime}_{-} = 0.$. The lower and upper momentum branches of the $y$-particle are
\begin{align}
\p_{-}(u) &= \arcsin(u),\qquad u\in (-1,1),\\
\p_{+}(u) &= \begin{cases}
\pi - \arcsin{(u)},&\quad u\in (0,1)\\
\arcsin{(-u)}-\pi,&\quad u\in (-1,0)
\end{cases},
\end{align}
with the corresponding derivatives
\begin{equation}
\p^{\prime}_{\pm}(u) = \pm \frac{1}{\ii (u\pm \ii 0)\sqrt{1-1/(u\pm \ii 0)^{2}}} =
\mp \frac{1}{\sqrt{1-u^{2}}},\qquad u \in (-1,1).
\end{equation}
The bare energies are $\e_{\pm}(u)$ and read
\begin{equation}
\e_{\pm}(u) = -2(\cos{(\p_{\pm}(u))}+\uu) = \pm 2\sqrt{1-u^{2}}-2\uu.
\end{equation}

\item
\underline{$M|uw$-strings}, which are bound states of $2M$ $u$-roots and $M$ $w$-roots, carrying charge $n_{M|uw}=2M$ and no
spin $m_{M|uw}=0$. An $M|uw$-string is parametrized by $u\in \mathbb{R}$ and comprises of rapidities
\begin{equation}
u_{\pm j} = u \pm (M+2-2j)\ii \uu,\qquad  w_{j} = u + (M+1-2j)\ii \uu,\qquad j=1,2,\ldots,M.
\end{equation}
To find the corresponding $y$-roots we assign $y_{+j}=x(u_{j})$ and $y_{-j}=1/x(u_{-j})$. The corresponding momenta
and energies are obtained by summing over all constituent $u$-roots (recall that $w$-roots carry no momentum)
\begin{align}
\p_{M|uw}(u) &= \sum_{j=1}^{M}\left(\p_{+}(u_{+j}) + \p_{-}(u_{-j})\right), \\
\e_{M|uw}(u) &= \sum_{j=1}^{M}\left(\e_{+}(u_{+j}) + \e_{-}(u_{-j})\right) = \e_{+}(u+M\ii \uu)+\e_{-}(u-M\ii \uu) \notag \\
& = 2\sqrt{1-(u+M\ii \uu)^{2}}+2\sqrt{1-(u-M\ii \uu)^{2}}-4M\uu,.
\label{eqn:pMvw_momenta_and_energy}
\end{align}
The derivative of their momenta satisfy $\p^{\prime}_{M|uw}<0$, reading explicitly
\begin{equation}
\p^{\prime}_{M|uw}(u) = \p^{\prime}_{+}(u+M\ii \uu) + \p^{\prime}_{-}(u-M\ii \uu) =
-\frac{1}{\sqrt{1-(u+M\ii \uu)^{2}}} - \frac{1}{\sqrt{1-(u-M\ii \uu)^{2}}}.
\end{equation}

\item
\underline{$M|w$-strings}, which are chargless ($n_{M|w}=0$) compounds made of $M$ $w$-roots, with spin $m_{M|w}=-M$.
They are parametrized by
$u \in \mathbb{R}$, and are of the form
\begin{equation}
w_{j} = u + (M+1-2j)\ii \uu,\quad j=1,2,\ldots M.
\end{equation}
\end{itemize}

The total charge $N$ and total spin $M$ in a Bethe eigenstate in terms of numbers of string excitations $N_{a}$
($a=\{\pm,M|uw,M|w\}$) is
\begin{equation}
N = N_{+} + N_{-} + \sum_{M=1}^{\infty}2M N_{M|uw},\qquad M = \sum_{M=1}^{\infty} M (N_{M|uw}+N_{M|w}).
\end{equation}

\paragraph*{Thermodynamic limit.}
In the thermodynamic limit, the solutions to Eqs.~\eqref{eqn:LiebWu} become densely distributed on the rapidity axis and
can be expressed in terms of particle densities $\rho_{a}$ which are defined as smooth densities of Bethe strings $u^{(a)}_{j}$,
\begin{equation}
\rho_{a}\left(u^{(a)}_{j}\right) = \lim_{L\to \infty}\frac{1}{L\left(u^{(a)}_{j+1}-u^{(a)}_{j}\right)}.
\end{equation}
Given a set of string solutions $\{u^{(a)}_{j}\}$, the unoccupied solutions to Bethe equations
$e^{\ii p_{a}(u)}\prod_{b}\prod_{j=1}^{N_{b}}S_{ab}(u-u^{(b)}_{j})=1$ are understood as the holes. The hole densities in
the thermodynamic limit are denoted by $\rhoh_{a}(u)$, while the total densities of a state are denoted
by $\rho^{t}_{a}(u)=\rho_{a}(u)+\rhoh_{a}(u)$.

In the Hubbard model, the densities of $M|uw$- and $M|w$-strings, denoted by $\rho_{M|uw}(u)$ and $\rho_{M|w}(u)$, respectively,
are supported on the whole real axis, $u\in \mathbb{R}$. On the other hand, the rapidity distributions
of the special $y$-particles are split into two separate densities $\rho_{\pm}(u)$ which are compactly supported
on the branch cut $\mathcal{I}$.
To this end, we define two types of integral transformations. First, we introduce the standard convolution as
\begin{equation}
(g\star h)(u,w)\equiv \int_{-\infty}^{\infty}\dd z\,g(u,z)h(z,w),
\end{equation}
taken with the convention that one drop $u$ and/or $w$ when $g$ and/or $h$ depend only on a single variable, and adopting
the implicit summation convention for convolution expressions of the form $g_{ab}\star h_{bc}$ and $g_{ab}\star g_{b}$, namely 
summing and integrating over the domain of a $b$-string. Since the $y$-particles' rapidity variable have a bounded integration
domain, i.e. $z\in \mathcal{I}$, we introduced a restricted convolution operation $\hstar$.
The densities of $y$-particles satisfy the sum rule,
\begin{equation}
1\hstar (\rho_{+}+\rho_{-}+\rhoh_{+}+\rhoh_{-}) = \frac{1}{2\pi}\hstar (p^{\prime}_{-}-p^{\prime}_{+}) = 1.
\end{equation}
Denoting $n_{M|uw}\equiv 1\star \rho_{M|uw}$, $n_{M|w}\equiv 1\star \rho_{M|w}$, and $n_{\pm}\equiv 1\hstar \rho_{\pm}$,
the electron charge and spin densities are expressed as
\begin{equation}
n = n_{+}+n_{-}+\sum_{M=1}^{\infty}2M\,n_{M|uw},\qquad m = \frac{1}{2}(n_{+}+n_{-})-\sum_{M=1}^{\infty}M\,n_{M|w}.
\end{equation}
Energy density of a macroscopic state is obtained by adding contributions of all energy-carrying particles,
\begin{equation}
\e = \int_{-1}^{1}\dd u \rho_{+}(u) \e_{+}(u) + \int_{-1}^{1}\dd u \rho_{-}(u)e_{-}(u) +
\sum_{M=1}^{\infty}\int_{-\infty}^{\infty} \dd u\, \e_{M|uw}(u)\rho_{M|uw}(u).
\end{equation}

Takahashi's equations for the densities in rapidity parametrization take the form
\begin{equation}
\begin{split}
\rho^{t}_{\pm} &= \mp\frac{\p^{\prime}_{\pm}}{2\pi} \mp K_{M}\star \left(\rho_{M|uw}+\rho_{M|w}\right),\\
\rho^{t}_{M|w} &= K_{M} \hstar \left(\rho_{+} + \rho_{-}\right) - K_{MN}\star \rho_{N|w},\\
\rho^{t}_{M|uw} &= -\frac{\p^{\prime}_{M|uw}}{2\pi} - K_{M} \hstar \left(\rho_{+}+\rho_{-}\right) - K_{MN}\star \rho_{N|uw}.
\end{split}
\label{eqn:BY_densities_canonical}
\end{equation}
The explicit form of integral kernels $K_{M}$ and $K_{MN}$ are given in section \ref{app:fusion_identities}.
Equations \eqref{eqn:BY_densities_canonical} can be, using fusion identities among scattering
kernels (cf. section \ref{app:fusion_identities}), decoupled in a quasi-local form
\begin{equation}
\begin{split}
\rho^{t}_{\pm} \mp s\star \left(\bar{\vartheta}_{1|uw} \rho^{t }_{1|uw}  + \bar{\vartheta}_{1|w} \rho^{t }_{1|w}\right)
&=  \mp \frac{1}{2\pi} \left(p^{\prime}_{\pm} - s   \star p^{\prime}_{1|uw}\right),\\
\left(\delta_{MN}\delta - I_{MN}\bar{\vartheta}_{N|uw}s\right)\star \rho^{t}_{N|uw}
&= \delta_{M,1}  s \hstar(\bar{\vartheta}_{-}\rho^{t}_{-} +
\bar{\vartheta}_{+} \rho^{t}_{+}),\\
\left(\delta_{MN}\delta - I_{MN}\bar{\vartheta}_{N|w}s\right)\star \rho^{t}_{N|w}
&=   \delta_{M,1}  s \hstar (\bar{\vartheta}_{-}\rho^{t}_{-} +
\bar{\vartheta}_{+} \rho^{t}_{+}),
\end{split}
\label{eqn:BY_densities_canonical}
\end{equation}
where $\delta_{MN}$ is the Kronecker delta, $\delta$ the Dirac delta, and the $I$ is the adjacency (incidence) matrix
for the model,
\begin{equation}
I_{MN} = \delta_{M,N-1} + \delta_{M,N+1}.
\label{eqn:adjacency_matrix}
\end{equation}

\paragraph*{Local statistical ensembles.}
Thermodynamic Bethe Ansatz method is based on expressing the free energy density of a local statistical ensemble
(a generalized Gibbs ensemble) as a set of coupled non-linear integral equations for the thermodynamic variables (e.g.
Fermi filling functions of the thermodynamic excitations). Generalized Gibbs ensembles are conventionally expressed in the form
\begin{equation}
\hat{\varrho}_{\rm GGE} \simeq \exp{\left(-\sum_{i}\mu_{i}\,\hat{Q}_{i}\right)},
\label{eqn:GGE_discrete}
\end{equation}
for a suitable (discrete) basis of local conserved quantities $\hat{Q}_{i}$ and the corresponding chemical potentials $\mu_{i}$. 
By accounting for the fact that particles' mode distributions $\rho_{a}(u)$ essentially contain the complete information about local 
correlations functions, it is convenient to consider as a starting point an analytic parametrization~\cite{IQC17}
\begin{equation}
\hat{\varrho}_{\rm GGE} \simeq \exp{\left(-\sum_{a}\int \dd u\,\mu_{a}(u)\hat{\rho}_{a}(u)\right)},
\label{eqn:GGE_analytic}
\end{equation}
where $\hat{\varrho}_{a}(u)$ correspond formally to a continuous family of local conserved operators whose eigenvalues coincide with
the particles' rapidity distributions, and $\mu_{a}())u$ are the chemical potentials pertaining to individual modes.
The partition sum $\mathcal{Z}_{\rm GGE}={\rm Tr}\,\hat{\varrho}_{\rm GGE}$ in the $L\to \infty$ limit is then evaluated with
a saddle-point integration using Yang--Yang approach~\cite{YY69}, where the entropy density per particle is the logarithm of
the number states occupying an infinitesimal rapidity interval $[u,u+\dd u]$ which (in models obeying
the Fermi statistics) takes a universal form
\begin{equation}
\mathfrak{s}_{a}(u) = \rho_{a}(u) \log \left(1+\frac{\rhoh_{a}(u)}{\rho_{a}(u)}\right) +
\rhoh_{a}(u)\log\left(1+\frac{\rho_{a}(u)}{\rhoh_{a}(u)}\right).
\label{eqn:Yang-Yang_entropy}
\end{equation}
A solution to the variational problem $\delta f[\rho_{a}]=0$, with
$f = \sum_{a}\int \dd u\,(\mu_{a}(u)\rho_{a}(u)-\mathfrak{s}_{a}(u))$,
yields \emph{canonical} TBA equations
\begin{equation}
\begin{split}
\log Y_{y} &= \mu_{y} + K_{N}\star \log \left(\frac{1+1/Y_{N|uw}}{1+1/Y_{N|w}}\right),\\
\log Y_{M|uw} &= \mu_{M|uw} + K_{MN}\star \log(1+1/Y_{N|uw})- K_{M}\hstar \log \left(\frac{1+1/Y_{-}}{1+1/Y_{+}}\right),\\
\log Y_{M|w} &= \mu_{M|w} + K_{MN}\star \log(1+1/Y_{N|w})- K_{M}\hstar \log \left(\frac{1+1/Y_{-}}{1+1/Y_{+}}\right),
\end{split}
\label{eqn:canonicalTBA}
\end{equation}
where the TBA $Y$-functions are as usual defined as ratios of hole and particle densities for each thermodynamic excitation
in the spectrum,
\begin{equation}
Y_{\pm} = \frac{\rhoh_{\pm}}{\rho_{\pm}},\quad Y_{M|uw} = \frac{\rhoh_{M|uw}}{\rho_{M|uw}},\quad
Y_{M|w} = \frac{\rhoh_{M|w}}{\rho_{M|w}}.
\label{eqn:Y-functions}
\end{equation}
The set of $Y$-functions is equivalent to the set of Fermi filling functions $\vartheta_{a}$, defined as
$\vartheta_{a}(u)=\rho_{a}/\rho^{t}_{a}$. For later purposes we moreover introduce the filling functions of the holes,
$\bar{\vartheta}_{a}(u)=1-\vartheta_{a}(u)$.

For instance, in canonical Gibbs equilibrium,
$\hat{\varrho}_{\rm Gibbs} \sim \exp{(-\beta \hat{H} + \mu_{\rm c}\hat{N} + \mu_{\rm s}\hat{S}^{z})}$,
the canonical source terms in terms of particles' bare energies $e_{a}(u)$ and chemical potentials for 
the electronic charge and spin read
\begin{equation}
\begin{split}
\mu_{y}(u) &= \beta\,e_{y}(u) - \mu_{\rm c} - \mu_{\rm s},\\
\mu_{M|uw}(u) &= \beta\,e_{M|uw}(u) - 2M\,\mu_{\rm c},\\
\mu_{M|w} &= 2\beta\,M\,\mu_{\rm s}.
\end{split}
\label{eqn:Gibbs_canonical_source}
\end{equation}

\begin{figure}[t]
\label{fig:Hubbard_lattice}
\centering

\begin{tikzpicture}[scale = 1]
\tikzstyle{M|w}  = [circle, minimum width=20pt, draw, fill=cyan!60, inner sep=0pt]
\tikzstyle{M|uw} = [circle,minimum width=20pt, draw, fill=red!60, inner sep=0pt]
\tikzstyle{y-}   = [circle, minimum width=20pt, draw, fill=yellow!60, inner sep=0pt]
\tikzstyle{y+}   = [circle, minimum width=20pt, draw, fill=yellow!60, inner sep=0pt]

\draw[very thin,color=lightgray] (0,0) grid (5.9,2);
\draw[very thin,color=lightgray] (0,0) grid (2,5.9);
\draw[->, >=latex, color=black] (0,0) -- (7,0) node[right] {\large $s$};
\draw[->, >=latex, color=black] (0,0) -- (0,7) node[left] {\large $a$};

\foreach \x in {1,...,4} \node[M|w] at (\x+1,1) {$\x|w$};
\foreach \y in {1,...,4} \node[M|uw] at (1,\y+1) {$\y|uw$};
\node[y-] at (1,1) {$-$};
\node[y+] at (2,2) {$+$};

\end{tikzpicture}
\caption{Y-system inscribed in the T-lattice: $Y$-functions $Y_{a,s}$ for the Hubbard model are assigned to bulk vertices of
the `T-hook lattice' of the associated $\mathfrak{su}(2|2)_{\rm c}$ Lie superalgebra~\cite{Gromov11,QF13,Cavaglia15}.
$Y$-functions are identified as $Y_{a,1}\equiv Y_{M|uw}$ for $a=M+1 \geq 2$ (red, vertical wing), $Y_{1,s}\equiv Y_{M|w}$ for
$s=2,3,\ldots$ (blue, horizontal wing), and the two-sheeted function $Y_{y}$ (yellow) which is assigned to $Y_{1,1}\equiv Y_{-}$ and 
the corner node $Y_{2,2}\equiv Y_{+}$.}
\end{figure}
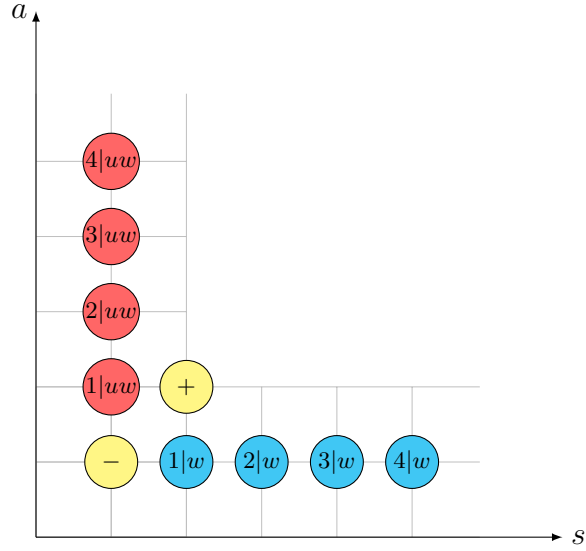

Using the fusion identities (cf. \ref{app:fusion_identities}), the above set of equations can be brought to an equivalent quasi-local 
form, reading explicitly
\begin{equation}
\begin{split}
\log Y_{\pm} - s \star \log\left(\frac{1+Y_{1|uw}}{1+Y_{1|w}}\right) &= \beta (e_{\pm} - s \star e_{1|uw}),\\
\log Y_{M|uw} - s \star I_{NM}  \log(1+Y_{N|uw}) &=
- \delta_{M1} s \hstar  \log \left(\frac{1+Y_{-}}{1+Y_{+}}\right) ,\\
\log Y_{M|w} -s \star  I_{MN}\log(1+Y_{N|w}) &=
- \delta_{M1} s  \hstar  \log\left(\frac{1+1/Y_{-}}{1+1/Y_{+}}\right),
\end{split}
\label{eqn:localTBA}
\end{equation}
supplemented with the asymptotic conditions
\begin{equation}
\lim_{M\to \infty} \frac{\log_{M|uw}}{M} = -2\mu_{\rm c},\quad \lim_{M\to \infty} \frac{\log Y_{M|w}}{M} = 2\mu_{\rm s}.
\end{equation}
By making the identifications $Y_{M|uw}\equiv Y_{M+1,1}$ and $Y_{M|w}\equiv Y_{1,M+1}$,
the Y-functions may be inscribed in the so-called $Y$-lattice (see e.g. \cite{AdSCFT_review,Cavaglia15}), as shown
on Fig.~\ref{fig:Hubbard_lattice}.

\subsection{Dressing of excitations and effective charges}
Excited states with respect to a reference macrosopic state (representing a many-body vacuum) are characterized in terms of
the particle-hole type of excitations and a background of non-excited modes (quantum numbers) which experience
a $\mathcal{O}(1/L)$ shift as a back-reaction to creating excitations. The difference between the rapidities
of excited and reference states induced by $N^{b}_{\rm ex}$ particle-type of excitations of type $b$ can be expressed as
\begin{equation}
\tilde{u}_{a,j} - u_{a,j} = \frac{1}{L}\sum_{b}\sum_{k=1}^{\rm N^{b}_{\rm ex}}
\frac{F_{ab}(u_{a,j},u_{b,k})}{\sigma_{a}\rho^{t}_{a}} + \mathcal{O}(L^{-2}),
\label{eqn:excited_states}
\end{equation}
while the hole-type excitations experience the same the shift of the opposite sign.
The shift functions $F_{ab}(u,w)$, describing the back-flow of non-excited rapidities, satisfy a closed set of integral equations
\begin{equation}
F_{ab}(u,w) = \frac{1}{2\pi}\phi_{ab}(u-w) - \sum_{c}\int \dd z K_{ab}(u-z)\sigma_{b}\vartheta_{b}(z)F_{bc}(z,w).
\label{eqn:F-function}
\end{equation}
In the thermodynamic limit, Eq.~\eqref{eqn:excited_states} can be expressed as an integral equation which
governs the dressing of bare quantities $q_{a}(u)$ (suppressing rapidity parameters)
\begin{equation}
q^{\rm dr}_{a} = q_{a} + q^{\prime}_{b}\,\vartheta_{b}\,\sigma_{b}\,\star F_{ba}.
\end{equation}
Differentiating this expression with respect to rapidity variable we find
\begin{equation}
q^{{\rm dr} \prime}_{a} = \Omega_{ab}\star q^{\prime}_{b},
\end{equation}
where the inverse of the dressing convolution kernel $\Omega$ explicitly reads
\begin{equation}
\left(\Omega^{-1}\right)_{ab}(u,w) = \delta_{ab}\,\delta(u-w) + K_{ab}(u-w)\,\vartheta_{b}(w)\,\sigma_{b}.
\end{equation}
Two special (but central) examples of the above transformation are the dressed energies $\edr_{a}$ and
dressed momenta $\pdr_{a}$, providing dispersion relations of the particle-hole excitations with respect to a reference macrostate.
The dressed velocities yield the group velocity of propagation and are given by
\begin{equation}
v^{\rm dr}_{a} = \frac{\dd \edr_{a}(\theta_{a}(u))}{\dd \pdr_{a}(u)} =
\frac{\edr^{\prime}_{a}(u)}{\pdr^{\prime}_{a}(u)}.
\end{equation}
Notice that particle velocities, which are given as a ratio of the derivatives of two dressed quantities, do not satisfy the
universal dressing equation. Moreover, it is worthwhile stressing that $\edr^{\prime}\neq (\e^{\prime}_{a})^{\rm dr}$ since
$[\partial_{u},\Omega_{ab}\star]\neq 0$. Hence, to avoid confusion, when applying the dressing transformation
to the bare charge densities we shall speak of the effective charges, that is
\begin{equation}
q^{\rm eff}_{a} = \Omega_{ab}\star q_{b}.
\end{equation}
The effective charges for the charge density $q_{i}$ can be alternatively obtained from the $Y$-functions of an
equilibrium state parametrized in the form of Eq.~\eqref{eqn:GGE_discrete} as
\begin{equation}
q^{\rm eff}_{a,i} = \partial_{\mu_{i}}\log Y_{a}.
\end{equation}

In the Hubbard model, the derivatives of the dress charges $q^{{\rm dr}\prime}_{a}$ are uniformly expressed as the solution to
the following system of integral equations,
\begin{equation}
\begin{split}
q^{{\rm dr}\prime}_{\pm} - s\star \left(\bar{\vartheta}_{1|uw}q^{{\rm dr}\prime}_{1|uw}  -
\bar{\vartheta}_{1|w} q^{{\rm dr}\prime}_{1|w}\right) &= q^{\prime}_{\pm} - s\star q^{\prime}_{1|uw},\\
\left(\delta_{MN}\delta - I_{MN}\bar{\vartheta}_{N|uw}s\right)\star q^{{\rm dr}\prime}_{N|uw} &=
\delta_{M,1}\,s \hstar (\bar{\vartheta}_{-}q^{\prime}_{-} - \bar{\vartheta}_{+}q^{\prime}_{+}),\\
\left(\delta_{MN}\delta - I_{MN}\bar{\vartheta}_{N|w}s\right)\star q^{{\rm dr}\prime}_{M|w}
&=- \delta_{M,1}\,s\hstar(\vartheta_{-}q^{\prime}_{-} - \vartheta_{+}q^{\prime}_{+}).
\end{split}
\label{eqn:Hubbard_dressing}
\end{equation}
For example, choosing $\pdr^{\prime}_{a} = \p^{{\rm dr}\prime}_{a}=2\pi\sigma_{a}\rho^{t}_{a}$
reduces Eq.~\eqref{eqn:Hubbard_dressing} to Eq.~\eqref{eqn:BY_densities_canonical}, while Eqs.~\eqref{eqn:localTBA} are retrieved by 
plugging in $\edr^{\prime}_{a} = \e^{{\rm dr}\prime}_{a}$.
Let us note that Eqs.~\eqref{eqn:Hubbard_dressing} comply with the morphology of the $Y$-system lattice, see Fig.~\ref{fig:Hubbard_lattice}.

The electron charge and spin have an exceptional role since they do not depend on rapidities variables.
Their effective values for the electron charge follow from the solution to
\begin{equation}
\begin{split}
n^{\rm eff}_{y} &= s\star (\bar{\vartheta}_{1|uw}n^{\rm eff}_{1|uw} - \bar{\vartheta}_{1|w}n^{\rm eff}_{1|w}),\\
n^{\rm eff}_{M|uw} &= I_{MN}s\star \bar{\vartheta}_{N|uw}n^{\rm eff}_{N|uw},\\
n^{\rm eff}_{M|w} &= 0,
\end{split}
\end{equation}
along with the large-$M$ asymptotic condition $\lim_{M\to \infty}n^{\rm eff}_{M|uw}=2M$. Similarly, for the effective spin we have
\begin{equation}
\begin{split}
m^{\rm eff}_{y} &= s\star (\bar{\vartheta}_{1|uw}m^{\rm eff}_{1|uw} - \bar{\vartheta}_{1|w}m^{\rm eff}_{1|w}),\\
m^{\rm eff}_{M|uw} &=0,\\
m^{\rm eff}_{M|w} &= I_{MN}s\star \bar{\vartheta}_{N|w}m^{\rm eff}_{N|w},
\end{split}
\end{equation}
with the asymptotics $\lim_{M\to \infty}m^{\rm eff}_{M|w}=M$. Therefore, only $M|uw$-strings and $y$-particles
yield non-vanishing effective charge, while $M|w$-stings and $y$-particle yield non-vanishing effective spin.
On the other hand, all types of particles (inducing the auxiliary ones) have non-zero effective energies $e^{\rm eff}_{a}$ in general.
The effective electron charges and spin can also be obtained from
\begin{equation}
n^{\rm eff}_{a} = \partial_{\mu_{\rm c}}\log Y_{a},\qquad
m^{\rm eff}_{a} = \partial_{\mu_{\rm s}}\log Y_{a}.
\end{equation}

\subsection{Scattering data and fusion identities}
\label{app:fusion_identities}

The elementary scattering amplitudes are
\begin{equation}
S_{M}(u) = \frac{u-M\ii \uu}{u+M\ii \uu},\qquad
S_{MN}(u)=S_{NM}(u)=S_{M+N}(u)S_{N-M}(u)\prod_{j=1}^{M-1}S_{N-M+2j}(u)^{2}.
\end{equation}
The scattering amplitudes for the regular $M|uw$-strings and $M|w$-strings are given by
\begin{equation}
S_{M|uw,N|uw}(u)=S_{MN}(u),\qquad S_{M|w,N|w}(u) = S^{-1}_{MN}(u),
\end{equation}
whereas the scattering amplitudes between $y$-roots and $M|uw$-strings or $M|w$-strings are $S_{M}(u)$.
The TBA integral kernels are defined as derivatives of the logarithmic scattering amplitudes,
\begin{align}
K_{M}(u) &= \frac{1}{2\pi \ii} \partial_{u}\log S_{M}(u) = \frac{1}{2\pi}\frac{2\uu M}{v^{2}+M^{2}\uu^{2}},\\
K_{MN}(u) &= \frac{1}{2\pi \ii}\partial_{u}\log S_{MN}(u) = K_{M+N}(u) + K_{N-M}(u) + 2\sum_{j=1}^{M-1}K_{N-M+2j}(u).
\end{align}
The kernels for $y$-particles are similarly given by
\begin{equation}
K_{\pm,a}(u) = \frac{1}{2\pi \ii}\partial_{u}\log S_{\pm,a}(u),
\end{equation}
for all types of particles $a$. Notice also that $K_{+a}=K_{-a}$.

Canonical TBA equations \eqref{eqn:canonicalTBA} can be cast in an equivalent quasi-local description by employing
the following fusion identities among the scattering kernels,
\begin{align}
K_{M} - s\star (K_{M-1}+K_{M+1}) &= \delta_{M,1} s,\qquad s(u) =\frac{1}{4\uu \cosh{(\tfrac{\pi u}{2\uu})}},\\
(K+1)^{-1}_{MN}(u) &\equiv \delta_{MN}\delta(u) - I_{MN}s(u),
\end{align}
with $K_{0}\equiv 0$. In addition, the latter satisfy
\begin{align}
(K+1)^{-1}_{MN}\star (K_{NQ}+\delta_{NQ}\delta) &= (K_{NQ}+\delta_{NQ}\delta)\star (K+1)^{-1}_{NM} = \delta_{MQ},\\
(K+1)^{-1}_{MN}\star K_{N} &= K_{N}\star (K+1)^{-1}_{NM} = \delta_{M1}s.
\label{eqn:standard_kernel_identities}
\end{align}
In the canonical (Gibbs) equilibrium, the action of $(K+1)^{-1}\star$ on the bare energies $e_{N|uw}$ yields
\begin{equation}
(K+1)^{-1}_{MN} \star e_{N|uw}=  \delta_{1M}s\hstar \left(e_{+}-e_{-}\right).
\end{equation}
Additionally, the terms which involve $\p^{\prime}_{M|uw}$ can be simplified using
\begin{equation}
(K+1)^{-1}_{NM}\star \p^{\prime}_{M|uw} = \delta_{M1} s\hstar \left(\p^{\prime}_{+}-\p^{\prime}_{-}\right).
\label{eqn:simpl}
\end{equation}

\section{Static covariance matrix}
\label{app:covariances}

To obtain the full static charge-current covariance matrix, i.e. overlap coefficients $\calO_{ij}$, we consider
\begin{equation}
\frac{\delta j_{i}}{\delta \mu_{j}}= \sum_{a}\int \dd u\,q_{a,i}(u)\delta_{\mu_{j}} j_{a}(u).
\end{equation}
To facilitate calculations, we introduce a compact vector notation for quantities depending solely on the mode labels $a$ and $u$,
namely $\bq=(q_{1}(u),q_{2}(u),\ldots)$, and similarly for other quantities.
Thus, expressing particle currents as
\begin{equation}
\bj = \left(\hat{\sigma}\hat{\vartheta}^{-1} + \hat{K}\right)^{-1}\frac{\be^{\prime}}{2\pi},
\end{equation}
readily yields
\begin{equation}
\delta_{\mu_{j}}\bj = -\left(\hat{\sigma}\hat{\vartheta}^{-1}\right)
\left(\hat{\sigma}\hat{\vartheta}+\hat{K}\right)^{-1}\frac{\be^{\prime}}{2\pi}.
\end{equation}
Employing $(\hat{\sigma}\hat{\vartheta}+\hat{K})^{-2}=\hat{\vartheta}^{2}\,\hat{\Omega}^{2}$,
and noticing that $\hat{\vartheta}^{-1}$ is a diagonal operator in the mode space, we find
\begin{align}
\delta_{\mu_{j}}j_{i} &= -\bq_{i}\cdot \hat{\sigma}(\delta_{\mu_{j}}\,\hat{\vartheta}^{-1})\,\hat{\vartheta}^{2}\,
\hat{\Omega}^{2}\,\frac{\be^{\prime}}{2\pi} = -\bq_{i}\cdot\hat{\sigma}(\delta_{\mu_{j}}\,\hat{\vartheta})\,\hat{\Omega}\,
\frac{\boldsymbol{\varepsilon}^{\prime}}{2\pi} = -\hat{\Omega}\,\bq_{i}\cdot \hat{\sigma}\,\hat{\vartheta}\,
\hat{\bar{\vartheta}}\,\frac{\boldsymbol{\varepsilon}^{\prime}}{2\pi}\bq^{\rm eff}_{j}
= \bq^{\rm eff}_{i}\cdot \hat{\rho}\,\hat{\bar{\vartheta}}\,\hat{v}^{\rm dr}\,\bq^{\rm eff}_{j}.
\end{align}
The mode kernel $O_{a}(u)$ for the charge--current correlator
$\calO_{ij}=\sum_{a}\int \dd u\,q^{\rm eff}_{a,i}(u)O_{a}(u)q^{\rm eff}_{a,j}(u)$ therefore takes the form
\begin{equation}
O_{a}(u) = \rho_{a}(u)\bar{\vartheta}_{a}(u)v^{\rm dr}_{a}(u).
\end{equation}

Alternatively, the static covariances can also be derived from the second derivatives of a functional
\begin{equation}
f_{g} = -\sum_{a}\int \frac{\dd u}{2\pi}\sigma_{a}g^{\prime}_{a}(u)\log\left(1+Y^{-1}_{a}(u)\right).
\end{equation}
By setting $g=\{\p,\e\}$ and calculating the gradients one obtains the well-known mode resolutions
\begin{equation}
\frac{\partial f_{\p}}{\partial \mu_{j}} = \sum_{a}\int \dd u\,q_{a,j}(u)\rho_{a}(u),\qquad
\frac{\partial f_{\e}}{\partial \mu_{j}} = \sum_{a}\int \dd u\,q_{a,j}(u)\rho_{a}(u)v^{\rm dr}_{a}(u).
\end{equation}
Note that $f_{\p}$ is the diagonal representation of the standard (generalized) free energy density $f$, which
can be readily deduced from combining Bethe--Yang equations for the densities, Yang--Yang entropy and TBA equations
for $\log Y_{a}$. The first derivatives of $\log Y_{a}$ with respect to chemical potentials give the effective charges,
\begin{equation}
\partial_{\mu_{j}}\log \bY = \hat{\Omega}\,\bq_{j} = \bq^{\rm eff}_{j},
\end{equation}
which readily implies
\begin{equation}
\frac{\partial f_{g}}{\partial \mu_{j}} = \sum_{a}\int \dd u\,g^{\prime}_{a}(u)\sigma_{a}\vartheta_{a}(u)q^{\rm eff}_{a,j}(u).
\end{equation}
From the identity
\begin{equation}
\partial_{\mu_{i}}\bq^{\rm eff}_{j} = -\hat{\Omega}\,\hat{K}(\partial_{\mu_{i}}\hat{\vartheta})\bq^{\rm eff}_{j},
\label{eqn:derivative_effective_charge}
\end{equation}
the second derivatives of $f_{g}$ take the form
\begin{equation}
\frac{\partial^{2}f_{g}}{\partial \mu_{i} \partial \mu_j} =
\sum_{a}\int \frac{\dd u}{2\pi}\,g^{\prime}_{a}\left((\partial_{\mu_{i}}\sigma_{a}\vartheta_{a})q^{\rm eff}_{a,j} + 
\sigma_{a}\vartheta_{a}(\partial_{\mu_{i}}q^{\rm eff}_{a,j})\right)
= \sum_{a}\int \frac{\dd u}{2\pi}\,(g^{\prime}_{a}(u))^{\rm eff}
\left(\partial_{\mu_{i}}\sigma_{a}\vartheta_{a}(u)\right) q^{\rm eff}_{a,j}.
\label{eqn:double_derivative_F}
\end{equation}
In the second line we have used
$\hat{1}-\hat{\sigma}\hat{\vartheta}(\hat{1}+\hat{K}\hat{\sigma}\hat{\vartheta})^{-1}\hat{K}=\hat{\Omega}$.
After expressing the derivatives of the filling functions $\vartheta_{a}$ as
\begin{equation}
\partial_{\mu_{i}}\vartheta_{a} = \frac{\partial \vartheta_{a}}{\partial Y_{a}}\frac{\partial Y_{a}}{\partial \log Y_{a}}
\frac{\partial \log Y_{a}}{\partial \mu_{i}} = -\vartheta_{a}\bar{\vartheta}_{a}q^{\rm eff}_{a,i},
\end{equation}
we obtain
\begin{equation}
\frac{\partial^{2}f_{g}}{\partial \mu_{i} \partial \mu_{j}} =
-\sum_{a}\int \frac{\dd u}{2\pi} q^{\rm eff}_{a,i}(u)(g^{\prime}_{a}(u))^{\rm eff}
\sigma_{a}\vartheta_{a}(u)\bar{\vartheta}_{a}(u)q^{\rm eff}_{a,j}(u),
\label{eqn:second_derivative_f}
\end{equation}
in turn implying
\begin{equation}
\calC_{ij} = -\frac{\partial^{2} f_{\p}}{\partial \mu_{i}\partial \mu_{j}},\qquad
\calO_{ij} = -\frac{\partial^{2} f_{\e}}{\partial \mu_{i} \partial \mu_{j}}.
\end{equation}
with the corresponding mode kernels
\begin{equation}
C_{a} = \frac{\pdr^{\prime}_{a}}{2\pi}\sigma_{a}\vartheta_{a}\bar{\vartheta}_{a}=\rho_{a}\bar{\vartheta}_{a},\qquad
O_{a} = \frac{\edr^{\prime}_{a}}{2\pi}\sigma_{a}\vartheta_{a}\bar{\vartheta}_{a}=\rho_{a}\bar{\vartheta}_{a}v^{\rm dr}_{a}.
\end{equation}

\paragraph*{Comment 1.}
The Bare quantities which take constant values (i.e. do not depend on rapidities) require careful considerations.
Considering spin of excitations as an example, the effective spin is determined via
\begin{equation}
(\delta_{ab}+K_{ab}\vartheta_{b})\star m^{\rm eff}_{b} = m_{a}.
\label{eqn:effective_spin_canonical}
\end{equation}
Since the bare spins $m_{a}$ of spin-carrying excitations are all non-zero, one would expect that the same holds automatically
also for the corresponding effective values. While this is in general the case, in the limit of vanishing chemical potential
$h\to 0$ the effective spin may tend to zero. The reason why this does not conflict with Eq.~\eqref{eqn:effective_spin_canonical} is 
an infinite summation over the particle content. Thus, in the presence of infinitely many types of excitations, the dressing 
transformation is of infinite dimension and demands an appropriate regularization.
To this end let us consider the Gibbs equilibrium state in $\beta \to 0$ limit where to the leading order in $h$ we
have $m^{\rm eff (0)}_{a}(h)=\tfrac{2}{3}(a+1)^{2}h+\mathcal{O}(h^{3})$, i.e. yields a results which is well-behaved in the
$h\to 0$ limit but diverges as $\sim a^{2}$ in the limit of large strings (large bare spin). The corresponding filling functions
on the other hand converge asymptotically for large $a$ as $\sim 1/a^{2}$, and read $\vartheta^{(0)}_{a}=1/(a+1)^{2}$.

A handy way to regularize the canonical form of Eq.~\eqref{eqn:effective_spin_canonical} is to use the fusion identities for
the scattering kernels to transform it in the quasi-local form. Using the fact that $(K_{ab}+\delta_{ab}\delta)\star m_{b}=0$,
Eq.~\eqref{eqn:effective_spin_canonical} is readily transformed into
\begin{equation}
m^{\rm eff}_{a} - s\star (m^{\rm eff}_{a-1}\bar{\vartheta}_{a-1}+ m^{\rm eff}_{a+1}\bar{\vartheta}_{a+1}) = 0.
\end{equation}
The erased source term is substituted with an appropriate large-$a$ asymptotic condition. For finite $h$ we should impose
$\lim_{a\to \infty}m^{\rm eff}_{a}=a$, which follows from the large-$a$ asymptotics of $\log Y_{a}$ in the canonical TBA equations.
In the $\beta \to 0$ limit and finite $h$, the Y-functions $\log Y^{(\infty)}_{a}(h)$ take the form
$1+\log Y^{(\infty)}_{a}(h)=\sinh^{2}{(h(a+1))}/\sinh^{2}{(h)}$. The $h\to 0$ limit is achieved by extrapolation.

\paragraph*{Comment 2.} One often deal with a situation when the values of effective charges $\bq^{\rm eff}_{j}$ exactly 
vanish, e.g. $m^{\rm eff}_{a}=0$ at zero chemical potential $h=0$. In such a case Eq.~\eqref{eqn:derivative_effective_charge} has 
to be regularized by considering small $h$ and only taking the limit $h\to 0$ at the end of computation, after first performing the 
infinite mode summation and a non-compact integration in Eq.~\eqref{eqn:second_derivative_f}. In the opposite case,
Eq.~\eqref{eqn:derivative_effective_charge} would imply vanishing derivatives $\partial_{h}s^{\rm eff}_{a}$ and thus an incorrect
result $\chi_{s}\equiv \calC_{ss}=0$.

\section{Drude weights from linearized hydrodynamics}
\label{app:linearized}

Drude weights can be defined as the variation of the equilibrium expectation values of the \emph{total} current\cite{ID17} with
respect to thermodynamic forces $\delta \mu_{j}$,
\begin{equation}
\calD^{(i,j)} = \frac{\beta}{2}\frac{\partial J_{i}}{\partial\,\delta \mu_{j}}|_{\delta \mu_{j}=0} =
\frac{\beta}{2}\sum_{a} \iint \dd \zeta\,\dd u\,q_{a,i}(u;\zeta)
\frac{\partial j_{a}(u;\zeta)}{\partial\,\delta \mu_{j}}|_{\delta \mu_{j}=0}.
\end{equation}
We imagine a bipartite initial state with a chemical potential drop $\delta \mu_{j}$ at the origin
(while keeping other chemical potential fixed).
The filling functions inside the light cone $\vartheta_{a}(u;\zeta) = \vartheta^{\rm L}_{a}(u) +
\Theta(v^{\rm dr}_{a}(u)-\zeta)\left(\vartheta^{\rm R}_{a}(u)-\vartheta^{\rm L}_{a}(u)\right)$,
with the left/right boundary conditions $\vartheta^{\rm L,R}_{a}$ which differ by amount $\mathcal{O}(\delta \mu_{j})$.
The corrections due to the difference of particle velocities only enter in the sub-leading order and can be disregarded.
On every ray $\zeta$, the particle current densities can be expressed as
\begin{equation}
\bj(\vartheta) = \left(\hat{\sigma}\hat{\vartheta} - \hat{\Xi}(\zeta) + \hat{K}\right)^{-1}\frac{\be}{2\pi},
\end{equation}
where $\hat{\Xi}=\hat{\Xi}(\zeta)$ is a \emph{diagonal} operator which involves the jump discontinuity,
\begin{equation}
\hat{\Xi} = \tfrac{1}{2}\left(1-2\Theta(\hat{v}^{\rm dr}-\zeta)\right)
\frac{\partial \hat{\vartheta}^{-1}}{\partial \mu_{j}}.
\end{equation}
Writing $\hat{A}=\hat{\sigma}\,\hat{\vartheta}\,\hat{\Omega}$, expanding the inverse
$(\hat{A}^{-1}+\hat{\Xi})^{-1}=\hat{A}+ \hat{\Xi}\hat{A}^{2}$, using the identities
\begin{equation}
\frac{\partial\,\bj(\zeta)}{\partial\,\delta \mu_{j}}|_{\delta \mu_{j}=0} =
-\hat{\Xi}(\zeta)\,\hat{\Omega}^{2}\,\frac{\be^{\prime}}{2\pi},\qquad
\frac{\partial\vartheta_{a}}{\partial \mu_{j}}=-\vartheta_{a}\bar{\vartheta}_{a}q^{\rm dr}_{a,j},
\end{equation}
and finally integrating over the light cone region,
\begin{equation}
\int_{\zeta=-\infty}^{\infty}\dd \zeta\,\tfrac{1}{2}\left(1-2\Theta(v^{\rm dr}_{a}(u)-\zeta)\right) = v^{\rm dr}_{a}(u),
\end{equation}
yields
\begin{equation}
D_{a} = \frac{\vartheta_{a}\bar{\vartheta}_{a}}{\rho^{t}_{a}}\left(\frac{\edr^{\prime}_{a}}{2\pi}\right)^{2}
=\rho_{a}\bar{\vartheta}_{a}(v^{\rm dr}_{a})^{2}.
\end{equation}
In the last equality we have used $\edr^{\prime}_{a}=2\pi \sigma_{a}\rho^{t}_{a}v^{\rm dr}_{a}$.

\section{On the detailed balance}
\label{app:detailed}

It was shown in \cite{SciPostPhys.1.2.015} that in the Lieb-Liniger model the dynamical density structure,
in the low-momentum limit $\kappa \to 0$, is determined by a single particle-hole excitation, 
\begin{align}\label{eq:FF}
& \mathcal{S}_{\hat{\rho}}(\kappa,\omega) = (2 \pi)^2  \Big| \left\langle \vartheta | \hat{\rho} | \vartheta , u - \tfrac{\kappa}{2p^{\prime}(u)} \to u +  \tfrac{\kappa}{2 p^{\prime}(u)} \right\rangle \Big|^2 \delta(\omega- \kappa v^{\rm dr}(u)) + O(\kappa^2),
\end{align}
where $\hat{\rho}$ is the conserved density operator and
$| \langle \vartheta | \hat{\rho} | \vartheta , u \to \tilde{u} \rangle |^2 $ represents a matrix element of the single particle-hole 
excitation $u \to \tilde{u}$ with energy $\omega$ and small momentum $\kappa$. Recall that the matrix element is
proportional to the available density of states for the particle-hole excitation, namely
\begin{equation}
| \langle \vartheta | \rho | \vartheta , u \to \tilde{u} \rangle |^2\sim \vartheta(u) (1-\vartheta(\tilde{u})).
\end{equation}
Therefore, expanding it to the first order in $\kappa$ around $\kappa=0$ we find
\begin{equation}
\vartheta (u - \tfrac{\kappa}{2p^{\prime}(u)}) \left(1-\vartheta (u + \tfrac{\kappa}{2p^{\prime}(u)}) \right) =
\vartheta(u)\left(1-\vartheta(u)\right) \left( 1 +  \tfrac{\kappa}{2} \tfrac{\partial_u \log(\vartheta^{-1} - 1)}{ p'(u)} \right) + \mathcal{O}(\kappa^2),
\end{equation}
implying a detailed balance condition in the order $\mathcal{O}(\kappa)$ of the form
\begin{equation}
 \frac{\mathcal{S}_{\hat{\rho}}(\kappa,-\omega)}{ \mathcal{S}_{\hat{\rho}}(\kappa,\omega)} = \left( 1 -   {\kappa}{} \frac{\partial_u \log(\vartheta^{-1} - 1) }{\partial_u p(u)} \right) \Big|_{v^{\rm dr}(u) \kappa = \omega}  \equiv e^{-\mathcal{F}(\kappa,\omega)} + O(\kappa^2),
\end{equation}
where the function $ \mathcal{F} (\kappa, \omega)$ is given by
\begin{equation}
\mathcal{F}(\kappa,\omega) = \kappa \frac{\partial \log (\vartheta^{-1}(u)-1)}{\partial p(u)}\Big|_{v^{\rm dr}(u) \kappa = \omega}
= \frac{\kappa}{\vartheta(u) ( \vartheta(u)- 1 )} \frac{\partial \vartheta (u)}{\partial p (u)} \Big|_{v^{\rm dr}(u) \kappa = \omega}.
\end{equation}

The recent results of \cite{DS17}, lifting standard hydrodynamics results \cite{Spohn_book} to the
generalized hydrodynamic theory \cite{BCNF16,CDY16}, imply that the dynamical structure factor for any conserved local
charge $\hat{q}$ is characterized by a single particle-hole contribution, with energy equal to $\kappa\,v^{\rm dr}(u)$.
Models with multiple particle require an additional summation over all particle types,
and formula \eqref{eq:FF} generalizes to
\begin{equation}
\mathcal{S}_{\hat{q}}(\kappa,\omega) = \sum_{a} \mathcal{S}_{\hat{q},a}(\kappa,\omega) = 
\sum_{a} (2\pi)^2 \Big| \left\langle \vartheta | \hat{q} | \vartheta , u_a-  \tfrac{\kappa}{2p_{a}^{\prime}(u)} \to u_a +  \tfrac{\kappa}{2p_{a}^{\prime}(u)} \right\rangle \Big|^2 \delta(\omega- \kappa v^{\rm dr}_a(u)) + O(\kappa^2)
\end{equation}
where, using Eq.~\eqref{eqn:resolution}, the zero momentum limit of the matrix element is given by
\begin{equation}
\lim_{\kappa \to 0} \Big| \left\langle \vartheta | \hat{q} | \vartheta , u_a-  \tfrac{\kappa}{2p_{a}^{\prime}(u)} \to u_a +  \tfrac{\kappa}{2p_{a}^{\prime}(u)} \right\rangle \Big|^2 = (2\pi)^{-1}\,\rho_a(u_a) (1- \vartheta_a(u_a)) (q_a^{\rm eff} )^2  .
\end{equation}
By repeating the logic of \cite{SciPostPhys.1.2.015}, we obtain the detailed balance expression for each particle type
\begin{equation}
\mathcal{S}_{\hat{q},a}(\kappa,-\omega)  =e^{-\mathcal{F}_a(\kappa,\omega)}  \mathcal{S}_{\hat{q},a}(\kappa,\omega), \qquad \mathcal{F}_a(\kappa,\omega)= \frac{\kappa}{\vartheta_a(u) ( \vartheta_a(u)- 1 )} \frac{\partial \vartheta_a (u)}{\partial p_a (u)} \Big|_{v^{\rm dr}_a(u) \kappa = \omega}.
\end{equation}

\section{Spin Drude weight in the anisotropic Heisenberg chain}
\label{app:Kohn}

We briefly revisit the exceptional case of ballistic spin transport in the anisotropic Heisenberg model at half filling,
a phenomenon which has attracted considerable attention in the past, see 
e.g.~\cite{Narozhny98,Peres99,Zotos99,AG02,HM03,FK03,BFKS05,Herbrych11,SPA11,Znidaric11,KBM12,Karrasch17}.
As explained and discussed in \cite{ID17}, the peculiar behaviour of the finite-temperature spin Drude weight at
$\mu_{\rm s}=0$ -- which is vanishing outside of the critical interval $|\Delta|<1$ where it exhibits a nowhere-continuous dependence
on interaction anisotropy $\Delta$ -- is directly related to the formation of an exceptional pair of excitations
which are charged under a hidden non-unitary conservation law~\cite{DeLuca16}.

In \cite{ID17}, the anomaly has been explained on the basis of symmetry properties of thermodynamic states under the
spin-reversal transformation, which rigorously confirmed the exact analytical high-temperature bound derived earlier
in \cite{PI13}. Given the that dressing of particle excitations is a property of the reference local equilibrium state,
the aim of this section is to shortly revisit this interesting case from the point of view of the linearized 
hydrodynamics. As explained below, this allows to reconcile our results with the previously obtained analytical resulting
derive from alternative approaches.

The finite-temperature spin Drude weights has been initially computed with aid of the Kohn formula, 
cf.~\cite{SS90,FK98,Zotos99,BFKS05}, expressing $\calD^{(s)}$ as the thermal average of the energy-level curvatures
with respect to applying twisted boundary conditions~\cite{Kohn64} (or equivalently, piercing a ring with a 
magnetic flux $\phi$). By resolving the $\mathcal{O}(L^{-2})$ corrections of the spectrum using TBA approach,
ref.\cite{Zotos99} finds an explicit expression for the spin Drude weight in the anisotropic Heisenberg model,
which in our notation reads
\begin{equation}
\calD^{(s)}_{\rm Kohn} = \frac{\beta}{2}\sum_{a}\int \dd u\, \rho^{t}_{a}(u)\vartheta_{a}(u)\bar{\vartheta}_{a}(u)
(\varepsilon^{\prime}_{a}(u))^{2}(\partial_{\phi}\gamma_{a}(u))^{2}.
\label{eqn:Zotos}
\end{equation}
where functions $\gamma_{a}$ describe $\mathcal{O}(1/L)$ shifts of Bethe roots due to the twisted boundaries. The
$\phi$-derivatives of $\gamma_{a}$ satisfy the following integral equations
\begin{equation}
2\pi\sigma_{a}\rho^{t}_{a}(\partial_{\phi}\,\gamma_{a}) =
m_{a} - K_{ab}\vartheta_{b}\sigma_{b}\star (\partial_{\phi}\,\gamma_{b})\rho^{t}_{b}.
\label{eqn:gamma_function}
\end{equation}
Identifying $m^{\rm eff}_{a} = 2\pi \rho^{t}_{a}\partial_{\phi}\gamma_{a}$ now allows us to interpret
Eq.~\eqref{eqn:gamma_function} as the dressing of spin, which readily implies that $\calD^{(s)}_{\rm Kohn}$ indeed agrees with
\begin{equation}
\calD^{(s)}=\sum_{a}\int \dd u\,(m^{\rm eff}_{a})^{2}D_{a}(u),\qquad
D_{a}(u)=\rho_{a}(u)(1-\vartheta_{a}(u))(v^{\rm dr}_{a}(u))^{2}.
\end{equation}

In the Heisenberg spin chain, the thermodynamic particle are magnons and bound states thereof, carrying a finite
amount of bare spin. Their effective spins with respect to a generic local equilibrium state are thus given by some
non-trivial finite quantities. Nevertheless, when approaching half filling $h\to 0$ (at finite temperatures) their effective spin 
exactly vanishes in the gapped and isotropic regimes ($|\Delta|\geq 1$).
In the gapless regime on the other hand, the number of distinct stable particles reduces to a finite set and
consequently the effective spin cannot entirely vanish for all excitations in the $h\to 0$ limit. Remarkably, it turn our that
the effective spin in a half-filled state exactly vanishes for all magnonic particles with an exception of a distinguished pair
of particles, in \cite{ID17} labelled by $a=\bullet,\circ$, which at $\Delta = \cos{(\pi m/\ell)}$ carry a finite
(temperature-independent) effective spin $m^{\rm eff}_{\bullet,\circ}=\pm \ell/2$. It thus follows from the mode resolution \eqref{eqn:resolution} that only this special pair of magnonic bound states contribute to spin Drude weight,
and since the latter have been shown to be the only excitations which transport non-unitary local conservation laws found
in \cite{Prosen11,PI13}, this automatically implies that the exact Mazur projection calculated in ref.~\cite{PI13} is complete.
The main conclusion of this section is thus that all three different definitions of spin Drude weights are equivalent.

\begin{figure}[b]
\includegraphics[width=0.6\hsize]{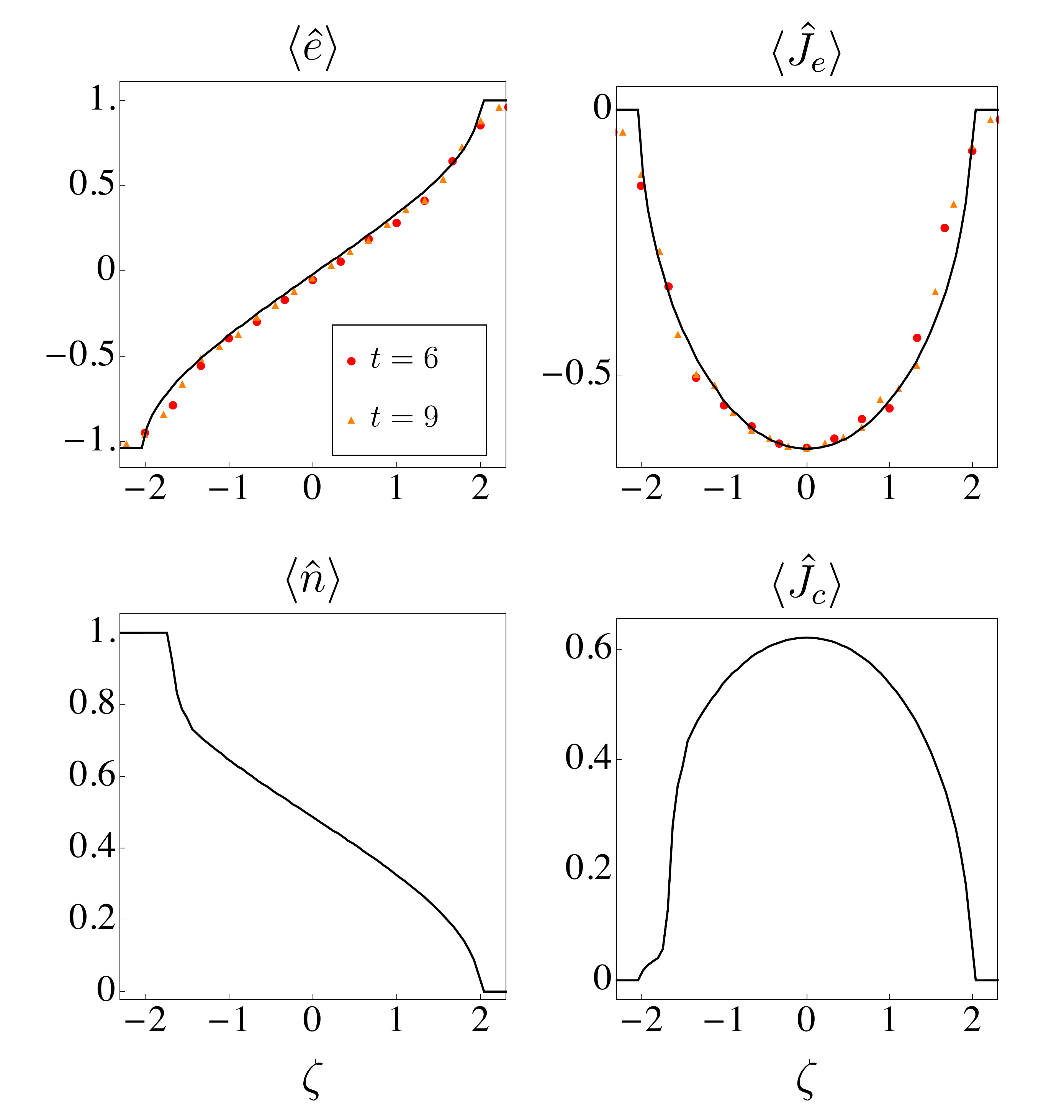}
\caption{Quasi-stationary profiles of energy density $e$ and charge density $n$, alongside the corresponding current profiles $J_{e}$ 
and $J_{n}$, computed from the hydrodynamic theory (black lines) and compared to tDRMG data reported in \cite{Karrasch17HUB}.}
\label{fig:profiles}
\end{figure}

\section{Hydrodynamic description of the Hubbard model}
\label{app:hydro_Hubbard}

We consider a non-equilibrium protocol such that the initial state is a tensor product of two different macroscopic states 
joined at the origin ($x=0$). This is usually referred to as the bi-partite protocol. In the long-time limit $t\to \infty$, the system 
is described locally by a quasi-stationary state which depends on the ray direction $\zeta = x/t$. The properties of
such states can be computed from the generalized hydrodynamic theory \cite{BCNF16,CDY16} which is formally a kinetic theory for the 
thermodynamic degrees of freedom of a system. In the Hubbard model, thermodynamic excitations comprise of spin-up electronic 
excitations (called the $y$-particles), the $M|uw$-strings (spinless electonic bound state) and the $M|w$-strings (spin-carrying 
chargeless bound states). Their root densities in position $x = \zeta t$ satisfy the usual hydrodynamic continuity equation
\begin{equation}
\begin{split}
\partial_t \rho_{y}(\zeta) &=  \partial_x \left(v^{\rm dr}_{y}(\zeta) \rho_{y}(\zeta)\right), \\
\partial_t \rho_{M|uw}(\zeta) &=  \partial_x \left(v^{\rm dr}_{M|uw}(\zeta) \rho_{M|uw}(\zeta)\right),  \qquad M= 1,2,\ldots,\\
\partial_t \rho_{M|w}(\zeta) &=  \partial_x \left(v^{\rm dr}_{M|w}(\zeta) \rho_{ M|w}(\zeta)\right),  \qquad M= 1,2,\ldots.
\end{split}
\label{eqn:continuity_SM}
\end{equation}
Solving the above set of equations automatically yields the correct distributions of the auxiliary particles $\rho_{M|w}$.
The solution to Eqs.~\eqref{eqn:continuity_SM} is given as usual in terms of the filling functions $\vartheta_{a}$
\begin{align}
\vartheta_{y}(u;\zeta) &=\Theta(\zeta -  v^{\rm dr}_{y}(u;\zeta)) \vartheta^{\rm R}_{y}(u) +
\Theta(-\zeta +  v^{\rm dr}_{y}(u;\zeta)) \vartheta^{\rm L}_{y}(u)  \label{eqn:continuity_solution} \\  
\vartheta_{M|uw,}(u;\zeta) &= \Theta(\zeta -  v^{\rm dr}_{M|uw}(u;\zeta)) \vartheta^{\rm R}_{M|uw}(u) +
\Theta(-\zeta +  v^{\rm dr}_{M|uw} (u;\zeta)) \vartheta^{\rm L}_{M|uw}(u),  \quad M= 1,2,\ldots \nonumber \\
\vartheta_{M|w}(u;\zeta) &= \Theta(\zeta -  v^{\rm dr}_{M|w} (u;\zeta)) \vartheta^{\rm R}_{M|w}(u) +
\Theta(-\zeta +  v^{\rm dr}_{M|w} (u;\zeta)) \vartheta^{\rm L}_{M|w}(u),  \quad M= 1,2,\ldots,
\label{eqn:continuity_solution2}
\end{align}
which completely determine the expectation values for local operators within the light cone emanating from the junction of the two
initial equilibrium states. We wish to point out that the $M|w$-strings, being the auxiliary degrees of freedom, do not supply
and dynamical information, but merely adapt their values to those of the physical ones, that is
the $y$ particles and the $M|uw$ strings. This implies that one can either obtain functions $\vartheta_{M|w}(u;\zeta)$ with 
Eq.~\eqref{eqn:continuity_solution2} or, alternatively, compute them via the TBA equations (suppressing dependence on $u$)
\begin{equation}
\log Y_{M|w}(\zeta) -s \star  I_{MN}\log(1+Y_{N|w}(\zeta)) =
- \delta_{M1} s  \hstar  \log\left(\frac{1+1/Y_{-}(\zeta)}{1+1/Y_{+}(\zeta)}\right),
\end{equation}
with $ Y_{y}(\zeta)= \vartheta_{y}^{-1}(\zeta)-1$ given by Eq.~\eqref{eqn:continuity_solution}. These two schemes yield
equivalent descriptions of the stationary state.
To demonstrate the above procedure, we present the profiles of the energy density and its current,
\begin{equation}
\hat{e}_{x} =  {-}\sum_{\sigma=\ua,\da}\Big(\hat{c}^{\dagger}_{x,\sigma}\hat{c}_{x+1,\sigma} +
\hat{c}^{\dagger}_{x+1,\sigma}\hat{c}_{x,\sigma}\Big) +
4\uu\left(\hat{n}_{x,\ua}-\tfrac{1}{2}\right)\left(\hat{n}_{x,\da}-\tfrac{1}{2}\right),
\end{equation}
\begin{equation}
\hat{J}_{e,x-1} - \hat{J}_{e,x}  = \ii [\hat{H}, \hat{e}_{x}],
\end{equation}
and moreover of the charge density operator 
\begin{equation}
\hat{n}_{x} =\sum_{\sigma=\ua,\da} \hat{c}^{\dagger}_{x,\sigma}\hat{c}_{x+1,\sigma},
\end{equation}
together with the charge current
\begin{equation}
\hat{J}_{c,x} =-\ii \sum_{\sigma=\ua,\da} \left(\hat{c}^{\dagger}_{x,\sigma}\hat{c}_{x+1,\sigma}  -
\hat{c}_{x,\sigma}\hat{c}^\dagger_{x+1,\sigma}    \right).
\end{equation}
We consider a thermal state with chemical potentials set to $(\beta, \mu, B)=(1, 0, 0)$ on one side, and the bare vacuum state
on the other (i.e. a thermal state with $\mu \to \infty$). Numerical results of our theoretical framework, obtained as the
solution to Eqs.~\eqref{eqn:continuity_solution} perfectly reproduce the DMRG data at late times as shown on Fig.~\ref{fig:profiles}.

\end{document}